\documentclass[10pt]{sig-alternate-05-2015}

\usepackage[pass]{geometry}

\usepackage{epsfig,endnotes}
\usepackage{epstopdf}
\usepackage{times}
\usepackage{color}
\usepackage{subfigure}
\usepackage{url}
\usepackage{graphicx}
\usepackage{soul}
\usepackage{xspace}
\usepackage[usenames,dvipsnames]{xcolor}
\usepackage{flushend}  %
\usepackage{etoolbox}
\usepackage{booktabs}  %
\usepackage{pbox}

\newif\ifcomment
\commentfalse

\newif\ifwatermark
\watermarkfalse

\ifwatermark
      \usepackage{draftwatermark}
      \SetWatermarkScale{.7}
      \SetWatermarkText{\shortstack{In Submission:\\Do Not Distribute}}
      \SetWatermarkColor[rgb]{1,0.88,0.88}
\fi

\ifcomment
    \newcounter{JBNumberOfComments}
    \stepcounter{JBNumberOfComments}
    \newcommand{\jbnote}[1]{\textcolor{magenta}{\small \bf [JB\#\arabic{JBNumberOfComments}\stepcounter{JBNumberOfComments}: #1]}}
    \newcounter{MVNumberOfComments}
    \stepcounter{MVNumberOfComments}
    \newcommand{\mvnote}[1]{\textcolor{blue}{\small \bf [MV\#\arabic{MVNumberOfComments}\stepcounter{MVNumberOfComments}: #1]}}
    \newcounter{DNNumberOfComments}
    \stepcounter{DNNumberOfComments}
    \newcommand{\dnnote}[1]{\textcolor{green}{\small \bf [DN\#\arabic{DNNumberOfComments}\stepcounter{DNNumberOfComments}: #1]}}

    \newcommand{\NOTE}[1]
    {
      {\footnotesize\it
        \begin{center}
          \begin{tabular}{|c|}
           \hline
            \parbox{0.85\columnwidth}{
              \medskip
              #1
              \medskip} \\
            \hline
          \end{tabular}
        \end{center}
        }    
    }
\else  
    \newcommand\jbnote[1]{}
    \newcommand\mvnote[1]{}
    \newcommand\dnnote[1]{}
    \newcommand\NOTE[1]{}
\fi

\newcommand\ourcomment[1]{}
\newcommand{\eg}{{e.g.,}\xspace}
\newcommand{\ie}{{\it i.e.,}\xspace}
\newcommand{\folder}{./fig}

\newcommand{\hh}{{HTTP/2}\xspace}

\newcommand{\h}{{HTTP/1.1}\xspace}

\newcommand{\eo}{{\small\sf{Eyeorg}}\xspace}

\newcommand{\cf}{{CrowdFlower}\xspace}
\newcommand{\lvc}{{\small\sf{LastVisualChange}}\xspace}
\newcommand{\fvc}{{\small\sf{FirstVisualChange}}\xspace}

\newcommand{\si}{{\small\sf{SpeedIndex}}\xspace}
\newcommand{\ol}{{\small\sf{OnLoad}}\xspace}

\newcommand{\proto}{{PROTO}\xspace}

\newcommand{\crawler}{{\small\sf{webpeg}}\xspace}
\newcommand{\uplt}{{\small\sf{UserPerceivedPLT}}\xspace}

\newcommand{\para}[1]{\vspace{0.1in}\noindent\textbf{#1}}

\newcommand{\paraittight}[1]{\vspace{0.05in}\noindent\textit{#1}}

\smallskip
\smallskip

\newcommand{\timeline}{timeline\xspace}
\newcommand{\Timeline}{Timeline\xspace}
\newcommand{\AB}{A/B\xspace}

\begin{document}

\CopyrightYear{2016} 
\setcopyright{acmcopyright}
\conferenceinfo{CoNEXT '16,}{December 12-15, 2016, Irvine, CA, USA}
\isbn{978-1-4503-4292-6/16/12}\acmPrice{\$15.00}
\doi{http://dx.doi.org/10.1145/2999572.2999590}

\date{}

\title{EYEORG: A Platform For Crowdsourcing Web Quality Of Experience Measurements}
\subtitle{\large\url{https://eyeorg.net/}}

\author{
\alignauthor 
  Matteo Varvello$^\dagger$,
  Jeremy Blackburn$^\dagger$,
  David Naylor$^\bullet$,
  Kostantina Papagiannaki$^\ddagger$\thanks{Work done while K. Papagiannaki was with Telef\'onica Research.}
\\
\affaddr{
  $^\dagger$Telef\'onica Research, $^\bullet$Carnegie Mellon University, $^\ddagger$Google Inc.}
}

\maketitle

\begin{abstract}
Tremendous effort has gone into the ongoing battle to make webpages load faster. This effort has culminated in new protocols (QUIC, SPDY, and HTTP/2) as well as novel content delivery mechanisms. In addition, companies like Google and SpeedCurve investigated how to measure ``page load time'' (PLT) in a way that captures \emph{human perception}. In this paper we present \eo~\cite{eyeorg_url}, a platform for crowdsourcing web quality of experience measurements. \eo overcomes the scaling and automation challenges of recruiting users and collecting consistent user-perceived quality measurements. We validate \eo's capabilities via a set of 100 \emph{trusted} participants. Next, we showcase its functionalities via three measurement campaigns, each involving 1,000 \emph{paid} participants, to 1)~study the quality of several PLT metrics, 2)~compare HTTP/1.1 and HTTP/2 performance, and 3)~assess the impact of online advertisements and ad blockers on user experience. We find that commonly used, and even novel and sophisticated PLT metrics fail to represent actual human perception of PLT, that the performance gains from HTTP/2 are imperceivable in some circumstances, and that not all ad blockers are created equal.
\end{abstract}



\begin{figure}[h]
  \centering 
    \subfigure[Timeline of \uplt responses.]{\includegraphics[width=0.9\columnwidth]{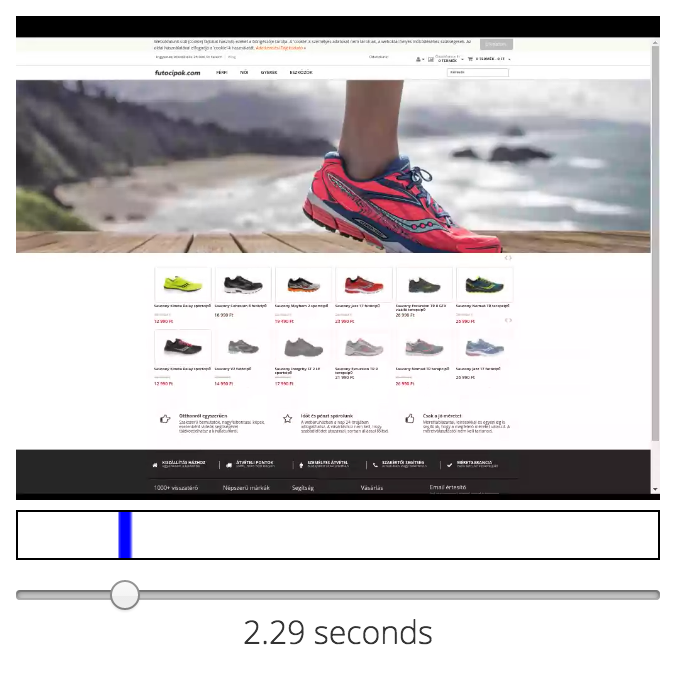}\label{fig:viz-tool-timeline}}
    \subfigure[Some sites exhibit multiple modes; here, some participants consider the site ``ready'' before the ads load.]{\includegraphics[width=\columnwidth]{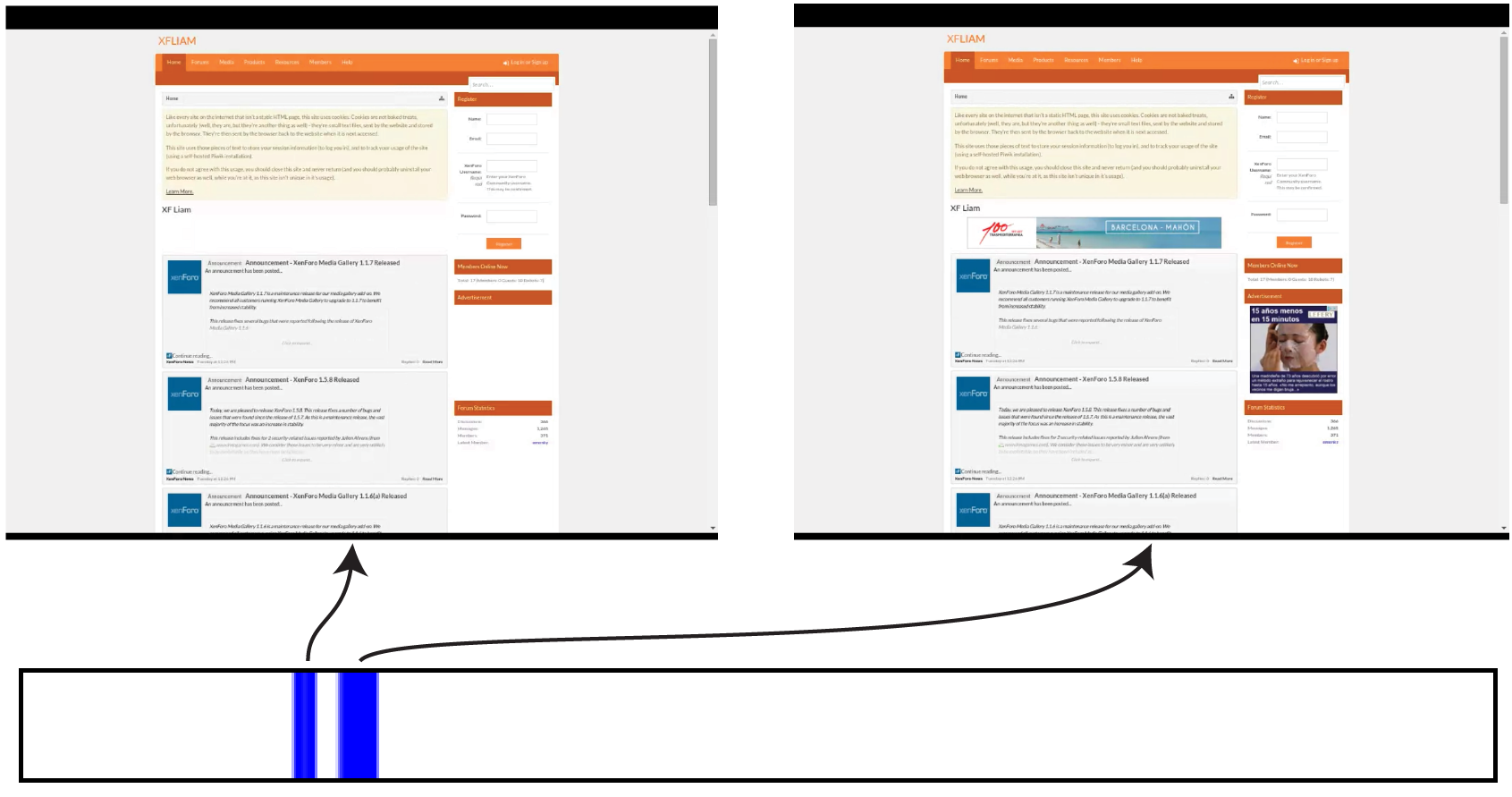}\label{fig:viz-tool-modes}}
  \vspace{-0.15in}
  \caption{Exploring responses with Eyeorg's visualization tools.}
  \dnnote{top figure might not be worth the space}
  \label{fig:viz-tool}
  \vspace{-0.15in}
\end{figure}

\section{Introduction}
\label{sec:intro}

Improving Quality of Experience (QoE) on the Web remains a hot topic. One obvious aspect of QoE, page load time (PLT), heavily impacts revenue for web-based companies~\cite{plt_revenue}. It is no surprise, then, that there are many recent efforts to decrease PLT from both industry  (\eg QUIC, SPDY, and HTTP/2) and academia~\cite{klotski, polaris, shandian}. To evaluate the effectiveness of such techniques, we need quantitative measures of their impact on QoE. Unfortunately, this is difficult because QoE is hard to define and even harder to measure.

Perhaps surprisingly, this is true even for PLT, which seems straightforward to
define and measure.  The typical PLT metric, the browser's ``onload'' event,
denotes when ``all of the objects in the document are in the DOM, and all the
images, scripts, links and sub-frames have finished
loading''~\cite{onload_mdn}.  \ol is not a perfect proxy for
\emph{user-perceived page load time} (\uplt) for two reasons: first, the user may
only care how quickly the \emph{visible} portion of the content loads (``above-the-fold''
render time), meaning \ol might be an overestimate. Second, scripts might continue
loading objects after \ol fires, in which case it may be an underestimate.  To
address this problem, new PLT metrics attempt to bridge the gap between \ol and
\uplt.  For example, \si\footnote{https://sites.google.com/a/webpagetest.org/docs/using-webpagetest/metrics/speed-index} defines PLT as the average time at which above-the-fold content is displayed.  Unfortunately, even this is not perfect: the user may consider the page ready to interact with even before some above-the-fold content (\eg ads) loads.

Ideally, we want a direct assessment of a user's experience on a webpage. This is difficult for the following reasons. First, a large number of people need to be recruited and surveyed. Second, participants need to be shown consistent views of a page loading. Third, we need to design experiments that lead to \emph{quantitative} results about a technique's impact on QoE.  In some studies~\cite{arapakis2014searchlatency, angeles2015physiologicallatency}, participants are shown a page load in person in a controlled environment and give qualitative feedback to interviewers. Clearly, this approach neither scales nor yields quantitative results.

\newpage
We take a different approach with \eo, a system for crowdsourcing Web QoE measurements. \eo allows researchers to test the impact
of changes to how a page is structured or delivered. \eo uses crowdsourced
participants to scale (we currently integrate with two popular services,
Microworkers and CrowdFlower); it shows \emph{videos} of pages loading to
provide a consistent experience to all participants, regardless of their
network connections and device configurations; and its surveys are designed to provide quantifiable feedback (\eg participants are asked to pick the point on a timeline when they consider a page loaded). With this approach, we maintain full control of experimental conditions and we can recruit any participant with a modern web browser, without requiring special hardware or software.

We present two kinds of results. First, we validate that \eo can produce good quality results; to do so, we recruit 100 \emph{trusted} participants and compare the quality of their answers with those of 100 \emph{paid} participants. 
We find that we can reliably identify unreliable participants, flagging about 20\% of the participants in our experiments as ``low performers'' whose results should be discarded. The speed with which crowdsourced participants can be recruited outweighs this overhead (\eg 1 hour rather than 10 days in this specific experiment). 

Next, we use \eo to 1) study how people perceive PLT, 2) compare \h and \hh performance, and 3) compare the QoE impact of three popular ad blockers: AdBlock, Ghostery and uBlock. For each measurement ``campaign,'' we recruited 1,000 paid participants to which we showed 5,000 page load videos (selected from 100 sites). Each campaign required 1.5 days to hit the 1,000 participants target. Our analysis shows that people find the \hh version of a website faster and that Ghostery outperforms its competitors. We also show that existing PLT metrics only partially match participants responses.  

To help understand the PLT data, we built a visualization tool that displays the \uplt responses as a timeline next to the video, as shown in Figure~\ref{fig:viz-tool-timeline}. Using this tool, we uncovered patterns in responses; for example, many videos have two modes, one for participants who consider the pages ``ready'' when the primary content is in place and one for those who wait for auxiliary content like ads (Figure~\ref{fig:viz-tool-modes}). The user-perceived PLT data we crowdsourced from 1,000 people is available at~\cite{eyeorg_url}.

\vspace{0.2cm}
\section{Related work}
\label{sec:related}

\subsection{Latency and Human Perception}

Bouch, Kuchinsky, and Bhatti performed one of the earliest in-depth studies to understand user perceptions and web page load time~\cite{bouch2000webqoe}.
They approached the problem from an HCI perspective by designing a controlled experiment where a small ($N = 30$) set of users were given an e-commerce related task to perform on the web.
They artificially introduced delay for each of the pages that the task required visiting and asked users to respond with a qualitative assessment (``high'', ``average'' , ``low'', and ``unacceptable'') of page performance. 
While they had many interesting findings, the one most relevant to \eo was that a mapping between subjective and objective PLT measurements was possible.
\eo takes the next step by greatly increasing the scale of responses as well as being a more general platform for conducting experiments.

Arapakis, Bai, and Cambazoglu investigated the impact of search engine response latency and user behavior in~\cite{arapakis2014searchlatency}.
Although their work was focused primarily on how latency affects user interaction and behavior for search engines, they provide some insight into how user perceived page load times might be measured.
In particular, in controlled experiments in a laboratory environment they asked participants to give an estimation of how long a page took to load.
They found that although individual responses did not match the controlled ground truth, the average of the estimates was quite close to the real values.
Although they used relatively coarse grained latency intervals (250~ms steps), their results indicate that aggregating human responses is a reasonable way to measure \uplt.

Egger et al.~\cite{Egger2012ICC} investigate the relationship between user perceived PLT and ``application'' PLT.\footnote{To the best of our knowledge, this is the browser reported ``onLoad'' time.}
To do so, they ask participants to mark the time at which they considered a page loaded.
The outcome of their study is that user-perceived PLT largely departs from application PLT: in most cases, users perceived a page as loading substantially faster than the application-reported PLT.
Their methodology differs significantly from ours, however.
In their study, participants browsed five different pages while link bandwidth and delay were manipulated.
In contrast, we show video captures of pages loading to thousands of study participants which might have subtle effects on their responses.
For example, our participants are able to precisely pinpoint the frame that they feel a page has loaded, fine-tuning their choice until satisfied.
We also measure a binary ``which is faster'' choice, which was not explored at all in~\cite{Egger2012ICC}.

\vspace{-0.1cm}
\subsection{Page Load Time}
\label{sec:related:PLT}
While the research community has focused on understanding the consequences of, and how to improve, page load time, it seems that understanding how to measure it is a mostly overlooked problem.
The greater ``web community,'' instead, has invested a lot into figuring out how to tell how fast their pages are being delivered to users.
These efforts range from blog posts discussing how to measure PLT~\cite{measuring.performance}, to ebooks explaining how to make your page load faster~\cite{bookofspeed}, to repositories of tools for measuring performance~\cite{repos}, to full on commercial measurement offerings~\cite{speedcurve}.
Even though there are serious business consequences associated with page performance, unfortunately, there is little science behind most of these efforts.

Regardless, it is worth exploring the efforts made in a bit more detail.
SpeedCurve~\cite{speedcurve} is a comprehensive suite of tools for web developers to gain a deeper understanding into page performance.
They provide instrumentation that extracts some of the metrics we use in this paper in addition to allowing the definition of custom metrics.
One interesting feature is that they provide developers a head-to-head benchmark comparing the developer's site with other, similar sites.
Ultimately, however, SpeedCurve blindly PLT metrics without any scientific justification that they are meaningful.
Determining the validity of computer-generated metrics is a primary goal of \eo.

In the scholarly community, ``onload'' has been the de facto PLT metric (e.g.,~\cite{klotski, polaris, shandian})---until recently.
Bocchi, De Cicco, and Rossi performed a preliminary study comparing different page load metrics~\cite{dario2016webqoe}.
They focused on evaluating two proposed metrics that are similar to SpeedIndex but are computationally less expensive.
Their most salient result is that correlating these metrics with real human perception is incredibly difficult, which is what \eo tries to address.

Nikravesh et al.~\cite{mobilyzer} measured web performance from 80 crowdsourced mobile devices using their library Mobilyzer.
They found a large degree of variability between PLT and what they call the	``page interaction time'' (similar to SpeedIndex).
The SpeedPerception\footnote{\url{http://speedperception.meteorapp.com/}} project has adopted a methodology similar to \eo's; in particular the idea of a human supplied response to a ``head-to-head'' matchup of page performance.
At the time of this writing, there are a few significant implementation differences: 1) SpeedPerception's \AB tests\footnote{SpeedPerception currently only supports \AB tests.} ask users to provide a decision on two \emph{different} sites and 2) the way that SpeedPerception implements \AB tests differs from \eo's (\S\ref{sec:design-experiments}).

\begin{figure*}[htb] 
\centering 
	\subfigure[\textbf{\Timeline Test.} Participants ``scrub'' the slider to the point where they consider the page ``ready to use.'']{\includegraphics[width=\columnwidth]{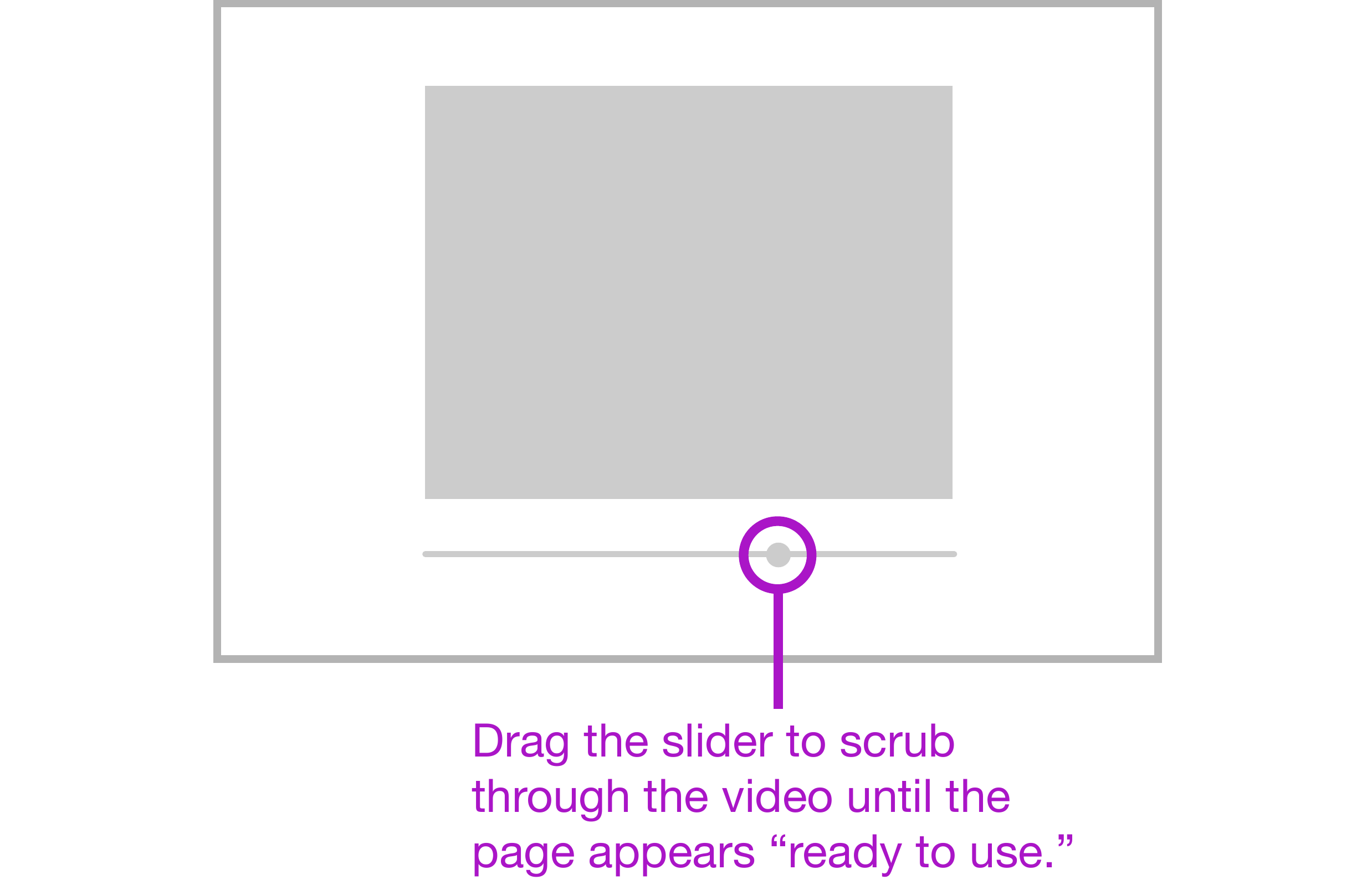}\label{fig:timeline}}
~
	\subfigure[\textbf{\AB Test.} Participants watch side-by-side page load videos and indicate which load is faster.]{\includegraphics[width=\columnwidth]{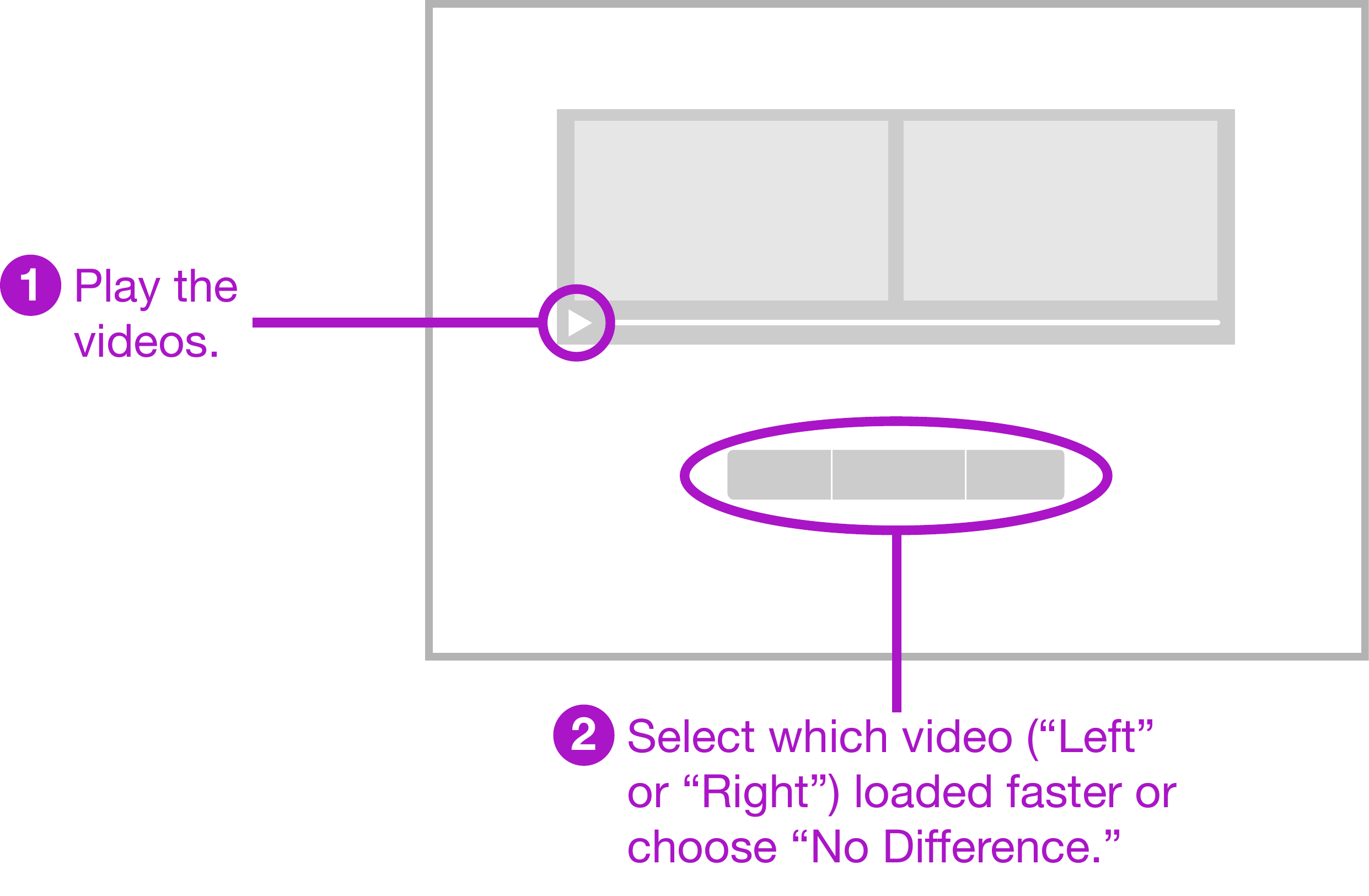}\label{fig:ab}}
\caption{\eo's experiment types.}
\label{fig:task} 
\end{figure*}

\section{Eyeorg}
\label{sec:design}

\dnnote{Give examples like inlining,
spriting, H2, shandian, klotski, polaris, etc.}

\eo~\cite{eyeorg_url} is a platform for crowdsourcing web quality of experience measurements; the goal of \eo is to measure improvements to user experience on the web. Both research and industry have expended great effort exploring different techniques for building and delivering web pages. Popular examples are content optimization techniques like ``inlining'' (putting CSS styles and JavaScript code  directly in HTML) and ``spriting'' (putting collection of images in a single image). More far-fetched solutions are shandian~\cite{Wang2016NSDI}, klotski~\cite{klotski}, and polaris~\cite{polaris} which aim at optimizing content delivery based on how it gets displayed from a browser.  

To understand when and how to use these techniques, we need to be able to \emph{measure their impact}.
Doing this in an automated and quantitative way is hard.
First, ``quality of experience'' is amorphous
and ill-defined---it is not obvious what to measure, let alone how.
Second, even aspects of Web experience that seem straightforward, like PLT, are difficult for machines to measure: ultimately, how \emph{humans} perceive performance is what matters.

\eo takes a new approach to quantifying quality of experience by allowing experimenters to directly test the impact of web page design or delivery techniques on real users in a controlled fashion at large scale. Experimenters can use \eo to answer questions like, ``Does changing the order in which objects load impact \uplt'', ``Does the presence of advertisements on a page negatively impact user experience?'', or ``Which demographics are more sensitive to PLT speedup?''. 

Experimenters can use \eo to draw responses from various classes of users, such as invited participants (\eg friends or colleagues), paid crowdsourced  workers, or general visitors to the \eo site. \eo provides tools to design experiments, recruit users, filter low-quality responses, and visualize results. The rationale behind \eo is that computer generated PLT metrics only partially capture how people perceive the web. The human perception data gathered by \eo can be used to augment and evaluate computer-generated PLT metrics.

\newpage  %
In the remainder of this section we describe the design and rationale behind \eo. In particular, we discuss how we address the following three challenges:

\begin{enumerate}

\item \textbf{How do we present page loads to participants?
(\S\ref{sec:design-overview})}\\Variations in load time caused by a particular
participant's device or network could mask variations due to the technique
being tested. Furthermore, experimenters might want to test the impact of
protocols or browser extensions that participants' browsers do not support.

\item \textbf{How do we ask questions about quality of experience (and get quantitative answers)?
(\S\ref{sec:design-experiments})}~~For example, how can participants indicate
to us when a page ``seems loaded'' (particularly non-technical participants)?

\item \textbf{How do we get lots of trustworthy responses?
(\S\ref{sec:design-crowdsourcing})}\\ Drawing meaningful conclusions requires a
large sample size; at the same time, recruiting participants not invested
in the experiments could yield careless, sloppy responses.

\end{enumerate}

\begin{figure*}[htb] 
\centering 
	\subfigure[Standard frame selection helper.]{\includegraphics[width=\columnwidth]{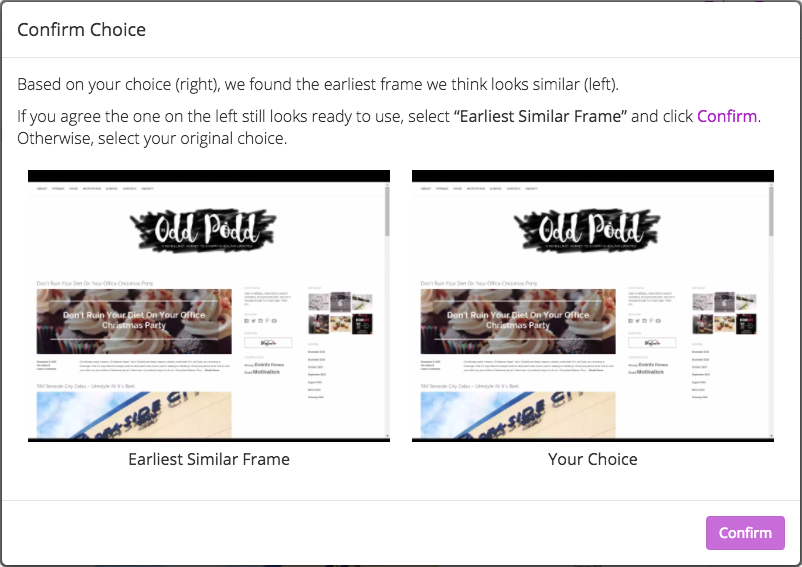}\label{fig:rewind-confirmation}}
	~~~
	\subfigure[Control frame selection helper.]{\includegraphics[width=\columnwidth]{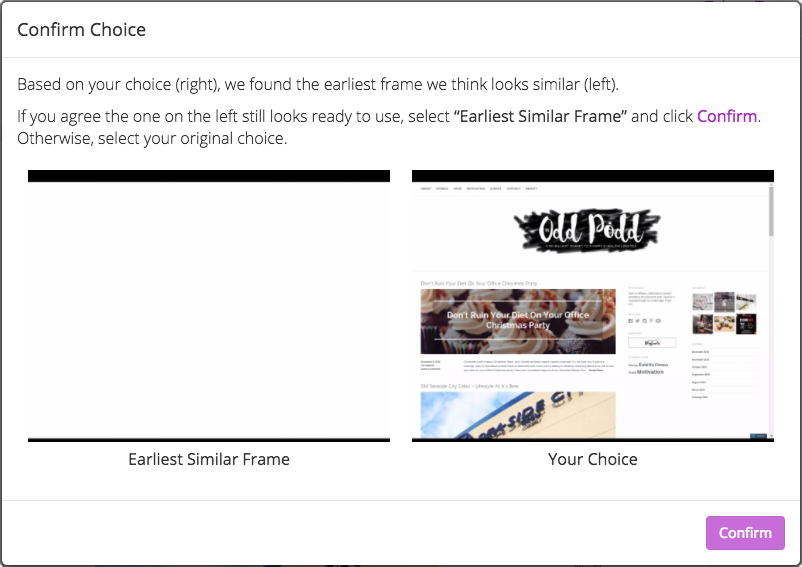}\label{fig:rewind-confirmation-control}} 		 			

\vspace{0.1cm}  %

\caption{Frame selection helper. \textnormal{\eo helps participants adjust
their selected PLT by giving them the option to choose the earliest visually
similar frame to the one they chose (left).  To ensure participants do not
blindly accept the ``rewind'' frame, as a control we occasionally show a
drastically different frame (right).}}
\label{fig:frame-selection-helper} 

\vspace{0.3cm}  %
\end{figure*}

\subsection{Providing a Controlled Experience}
\label{sec:design-overview}

Apart from the engineering challenge to build \eo, a more fundamental challenge lies in dealing with real people.
Testing participants' reactions to a web experience ``in the wild'' and at scale is difficult for several reasons.
First, differences between participants' browsers could impact results.
Second, in many cases the outcome we want to test (like PLT) is dependent on network
conditions, but we have no control over the quality of participants' network access.
Third, experimenters may want to test the impact of protocols (\eg HTTP/2 or SPDY) that
some participants' browsers may not support, or browser extensions (\eg AdBlockers) that some participants may not have installed.
In short, we want to guarantee the same experience for each participant.

To ensure that all participants base their responses on identical
experiences, \eo uses \emph{video}. This allows us to fully control what participants see---we pick the browser, network conditions, protocols, and
plugins---regardless of individual participants' configurations.

We built a tool, \crawler, to record videos of web pages loading.
Pages are loaded with Chrome because: 1)~unlike headless browsers like PhantomJS\footnote{http://phantomjs.org/}, Chrome provides rapid (and optimized) support for new technologies like SPDY and HTTP/2 and 2)~Chrome offers better support for instrumentation than Firefox or Safari.  We use Xvfb\footnote{https://www.x.org/archive/X11R7.7/doc/man/man1/Xvfb.1.xhtml} (the X virtual frame buffer) so we can run \crawler on machines without displays (\eg EC2 instances) and we capture videos in the webm format\footnote{https://www.webmproject.org/} (which offers small file sizes) using ffmpeg.\footnote{https://ffmpeg.org/}  

We designed \crawler to be highly customizable. We use Chrome's command line options to control things like protocol (\h or \hh) and appearance (kiosk mode) and Chrome's remote debugging protocol\footnote{https://developer.chrome.com/devtools/docs/debugger-protocol} to enable device and network emulation. The remote debugging interface also gives us detailed information about the page load (as an HTTP Archive, or HAR), including when each object loaded, which protocol was used, and when the onload event fired.
An alternative to using Chrome's remote debugging interface would have been to use the Navigation Timing API.\footnote{\url{https://developer.mozilla.org/en-US/docs/Web/API/Navigation_timing_API}}
However, the Navigation Timing API is designed to be accessed by web ``applications'' themselves via JavaScript.
To avoid \emph{any} impact our methodology might have on performance, we thus chose to use the asynchronous debugging protocol. In the future, we will explore performant uses of the NT API when expanding Eyeorg to other browsers.
Finally, \crawler  directly modifies Chrome's preference file to  enable/disable extensions and turn off distracting messages like ``Would you like to translate this page?''. 

To capture videos, the experimenter supplies a list of URLs, how many loads to perform per site, and how many seconds to record after onload (since there is no automatic way for \crawler to know when the page has finished loading---if there were, \eo would be unnecessary!). A fresh browser state is ensured by deleting Chrome's local state after each load. As in~\cite{Varvello2016}, before testing a new website, an initial ``primer'' load is performed. The primer ensures that all needed DNS records are cached at the ISP's DNS resolver before the first real trial (to prevent a cache miss from skewing the load time results). Local content and DNS caches are disabled and each request carries a cache-control header instructing network caches not to respond.

\vspace{0.6cm}  %

\subsection{Designing Experiments with Eyeorg}
\label{sec:design-experiments}

Despite its subjective nature, we would like to run
quantitative experiments measuring how QoE is impacted by various content or
delivery optimizations. 
To demonstrate how \eo can be used to do this, we
picked one aspect of user experience, page load time, and designed two initial types of experiments to study it: \timeline and \AB. \mvnote{here we should give hints how these experiments can be used for alternative goals}

\para{\Timeline}~~In a \timeline experiment, we present the participant with
a page load video and ask them to choose the point when they consider the
page ``ready to use.'' The naive way to design this test is to let the
participant play the video and ask them to pause it when the page is loaded;
however, since they may not be familiar with the page, they will likely wait
until they are sure no more content will load, overshooting and pausing too
late. Instead, we disable the standard HTML5 video controls and present
participants with a slider they use to ``scrub'' through the video
(Figure~\ref{fig:timeline}). Even with the slider, in early tests we observed
that (both trusted and paid) participants still overshot, choosing times well
beyond onload.  We suspect this is due to a combination of two reasons. First,
browsers often only preload a small portion of the video at a time, so when
participants seek quickly, they may see a blank screen and assume the
\emph{page in the video} has not loaded yet, when in reality the \emph{video}
has not loaded yet. By the time the video catches up, they have overshot. To
address this, for \timeline tests we force the browser to preload the entire
video before the test begins, only then enabling the slider.  Second, to
correct for simple sloppiness, we add one final step: after the participant
chooses a time, we show them the frame they chose and the earliest similar
frame (no more than 1\% different in a pixel-by-pixel comparison (Figure~\ref{fig:rewind-confirmation}). The participant can either accept our
``rewind frame'' or stick with their original choice.

We use \timeline tests to compare participants' perceived load time with other metrics like onload and SpeedIndex. Since these metrics are frequently used to evaluate techniques for improving PLT, it is important to understand how closely they match user-perceived load time.  Our tests use page load videos for a sample of 100 of the Alexa top 1M sites that fully support HTTP/2~\cite{h2_page}. For each experiment configuration, we repeat each load five times and use the video with the median onload time. Results are described in \S\ref{sec:plt-results}.

\para{\AB}~~The \timeline test is tricky; ``ready to use'' is subjective
and participants are not always sure how to decide what time to pick. However,
for some experiments, it is not important to choose precisely \emph{when} a
page is loaded; it may be useful simply to indicate which of two page loads is
\emph{faster}. For these cases, \eo uses \AB experiments. Participants watch
two page load videos simultaneously and pick which loaded faster or ``No
Difference'' (see Figure~\ref{fig:ab}).  Video pairs are shown in a random
order (i.e., ``A'' is not always on the left and ``B'' is not always on the right). There is no guarantee that two videos in a browser stay perfectly synchronized; for
instance, lost packets might momentarily stall one video while the other
continues playing. To ensure the videos stay synchronized, we splice them into
a single video file. If playback stalls, both sides are affected equally.

\begin{table*}[t]
\centering
\begin{tabular}{rcccccc}
\toprule

           & \pbox{20cm} {\bf Type} & {\bf Participants} & {\bf Male/Female} & {\bf Duration} & {\bf Cost} & {\bf \# Sites} \\
\midrule
\textit{\textbf{Validation}}    & & & & & & \\
\midrule
{\bf PLT} & \timeline    & Paid      & 76/24    & 1 hour    &  \$12  &  20  \\
{\bf PLT} & \timeline    & Trusted   & 79/21    & 10 days   &  -     &  20  \\
{\bf H1-H2}       & \AB  & Paid      & 77/23    & 1 hour    &  \$12  &  20  \\
{\bf H1-H2}       & \AB  & Trusted   & 84/16    & 10 days   &  -     &  20  \\
\bottomrule

\textit{\textbf{Final}}     & & & \\
\midrule
{\bf PLT} & \timeline    & Paid      & 685/315    & 1.5 days    &  \$120  &  100  \\
{\bf H1-H2}       & \AB  & Paid      & 697/303    & 1.5 days    &  \$120  &  100  \\
{\bf ADS}       & \AB    & Paid      & 716/284      & 1.5 days    &  \$120  &  100  \\
\midrule
\end{tabular}
~
\begin{tabular}{ccc}
\toprule

{\bf Engagement} & {\bf Soft} & {\bf Control} \\
\midrule
& & \\
\midrule
16 &  2 &  7 \\
10 &  - &  1\\
9 &  5 &  2\\
1 &  2 &  1\\
\bottomrule

& & \\
\midrule
151 &  45 &  54 \\
98  &  56 &  82 \\
128 &  34 &  57\\
\midrule
\end{tabular}

\caption{Summary of data collected. \textnormal{Engagement, Soft, and Control indicate the number of participants filtered out by the techniques described in Section~\ref{sec:cleaning}.}}
\label{tab:summary} 
\vspace{0.15cm}
\end{table*}

We demonstrate \AB test's versatility with two experiments: 1)~comparing the speed of HTTP/1.1 vs.~HTTP/2, and 2)~comparing the impact of ad blockers on PLT.  For HTTP/1.1 vs.~HTTP/2, we capture videos of the same 100 websites above while loading over HTTP/1.1 and HTTP/2. For the ad blocker analysis, we use data from~\cite{h2_page} to identify 10,000 websites that display ads. From this list, we sample 100 websites and make videos of page loads through Chrome using one of three popular ad block extensions: AdBlock, uBlock, and Ghostery. In this case we do not control the protocol, which means Chrome will default to HTTP/2 if the target website supports it. For each experiment configuration (\eg ``\hh'' or ``Ghostery''), we repeat each load five times and keep the video with the median onload time. Results are described in \S\ref{sec:plt-results}.

\subsection{Crowdsourcing \& Response Validation}
\label{sec:design-crowdsourcing}

For large sample sizes, \eo turns to crowdsourcing. While many crowdsourcing
services offer built-in tools for creating and hosting tests, we chose to build
our own infrastructure. First, the built-in facilities tend to be limited and are not flexible enough to
implement our \timeline and \AB tests. Second, we did not want to tie \eo to
one particular crowdsourcing service; we wanted the ability to draw from a
larger, more diverse participant pool, including those not registered as crowd
workers, like friends and colleagues.
\eo currently supports
Microworkers\footnote{https://microworkers.com} and CrowdFlower\footnote{https://www.crowdflower.com} and can easily be extended to
integrate with other platforms.

Using crowdsourced participants poses a challenge: we need to verify that responses are meaningful, \eg that paid participants do not just blindly click through our tests to get paid (there are indications that workers from well known services like Mechanical Turk\footnote{https://www.mturk.com/mturk/welcome} do not always perform as well as desired~\cite{crowdflower.shits.on.mturk}).

While there is no standardized methodology to determine the quality of a crowd worker, we draw on existing research in the field to detect unreliable responses (using a combination of test-time mechanisms and after-the-fact filtering).

\para{Hard Rules}~~Participants are given a set of clear instructions
for each test, defining a set of \emph{hard rules} they must follow for the system to allow them to complete the test. For example, in \AB tests,
participants must choose ``Left,'' ``Right,'' or ``No Difference'' in order to move to the next video.  Because we are interested in \emph{human perception}, we also use Google's ``I'm not a robot'' service to verify ``humanness'' before participants take tests.

\para{Soft Rules}~~These are rules that, while not strictly enforced by the system, can reflect the conscientiousness of the workers~\cite{Rainer2015}. \eo does not force participants to watch a video before answering our questions, but failure to do so is a sign of low quality.

\para{Engagement}~~ High quality workers tend to be ``active.'' While the total amount of time a worker spends on a task is a rough indication of the quality of their work~\cite{Rzeszotarski2011}, it is not enough.  On the one hand, a short time might indicate an unengaged worker whose only goal is finishing the task as fast as possible; on the other hand, a long time might be due to a distracted worker.   We track how long
participants spend reading the instructions, how much time they spend on each video, how many times they play it (if at all), how much of the video they watch, how often they seek, and whether they switch to different tabs or windows during a test. We also capture information about the participant's system, like browser, OS, and how large the video was on their screen.

\para{Control Questions}~~ A common technique in crowdsourcing experiments is to randomly insert \emph{control questions}---questions to which the answer is known~\cite{Hossfeld2014}.  For \timeline tests, we want to verify that participants do not blindly accept the rewind frame we suggest, so we occasionally suggest a nearly-blank rewind frame (Figure~\ref{fig:rewind-confirmation-control}) and check that the participant continues with their original choice. For \AB tests, we occasionally show two copies of the same video with one side artificially delayed by three seconds and check that the participant picks the non-delayed side.

\para{Wisdom of the Crowd}~~ A classic problem in crowdsourcing is the lack of ground truth, meaning the quality of responses is difficult to evaluate. However, if most participants are reasonably good, ground truth can be built using the average or majority vote of all responses~\cite{NIPS2010_0577, blackburn2014stfunoob}. Participants whose responses deviate wildly from this pseudo-ground truth can be dropped.  We also leverage the crowd to improve the system: participants can report broken videos; a video flagged by 5 different workers is automatically banned and manually inspected by our team.

\newcommand{\takeaway}[1]{\smallskip \noindent \textit{\textbf{Takeaway:}~#1}}

\vspace{0.3cm}
\section{Validating Eyeorg's Results}
\label{sec:validation}

In this section, we validate the quality of the crowdsourced responses that \eo
collects.  We introduce the data sets we have collected
(\S\ref{sec:validation:datasets}) and present an analysis of various techniques
for throwing away unreliable responses (\S\ref{sec:validation:results}),
which leads us to our final ``cleaning'' procedure  (\S\ref{sec:cleaning}).

\newpage 
\subsection{Data Sets}
\label{sec:validation:datasets}
Before exploring methods for detecting low quality workers (\S\ref{sec:design-crowdsourcing}), we define a ``pseudo-ground truth'' baseline using responses from a set of trusted participants. We evaluate the effectiveness of different filtering techniques by comparing filtered paid participant responses against these baseline trusted participant responses.

To gather the baseline, we ran two small-scale campaigns, one \AB\footnote{The \AB campaign is HTTP/1.1 vs HTTP/2.} and one \timeline, with 20 videos each. For each campaign, we recruited 100 \emph{paid} participants on \cf and 100 \emph{trusted} participants via email and social media. On \cf, we requested only workers that are ``historically trustworthy,''
which comes at the cost of a longer recruitment time. For the trusted participants, we carefully selected friends and colleagues who promised full commitment to the task. We asked each participant to watch six videos (\ie we served 600 videos in total per experiment and each video has, on average, responses from 30 different participants). 

The top of Table~\ref{tab:summary} summarizes the validation data set. For both \timeline and \AB campaigns, it took about one hour to recruit paid participants (cost: \$12) compared to 10 days for the trusted participants (at no cost). Both participant sets exhibit a roughly 75/25\% male/female gender split. Paid participants are located in 30~countries (Venezuela being the most popular); trusted participants came from 12~countries (the U.S. being the most popular).

\begin{figure*}[htb] 
\centering 
	\subfigure[Time spent on site]{\psfig{figure=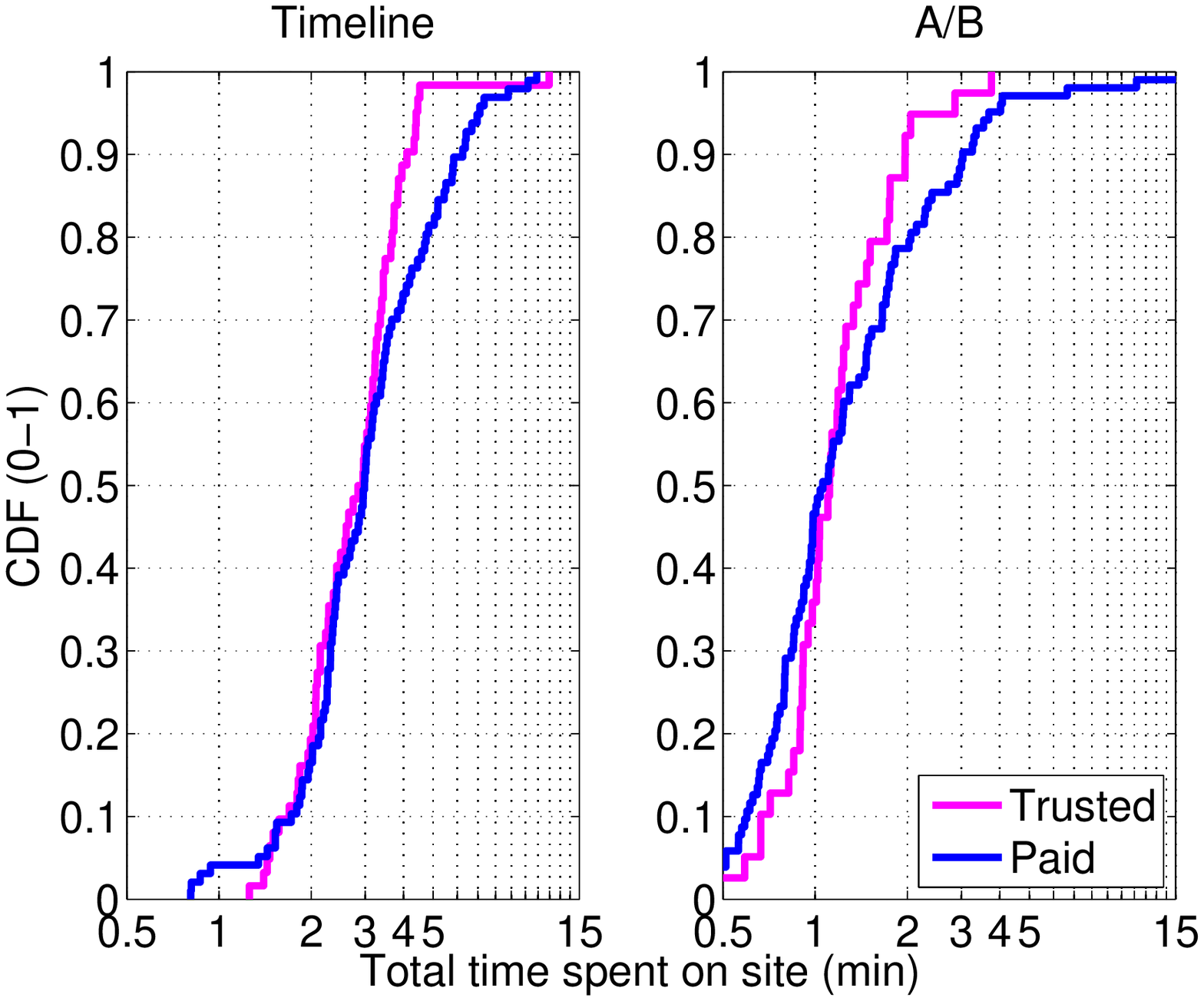,width=2.1in} 
	\label{fig:validation-a}}
	\subfigure[Number of actions on site]{\psfig{figure=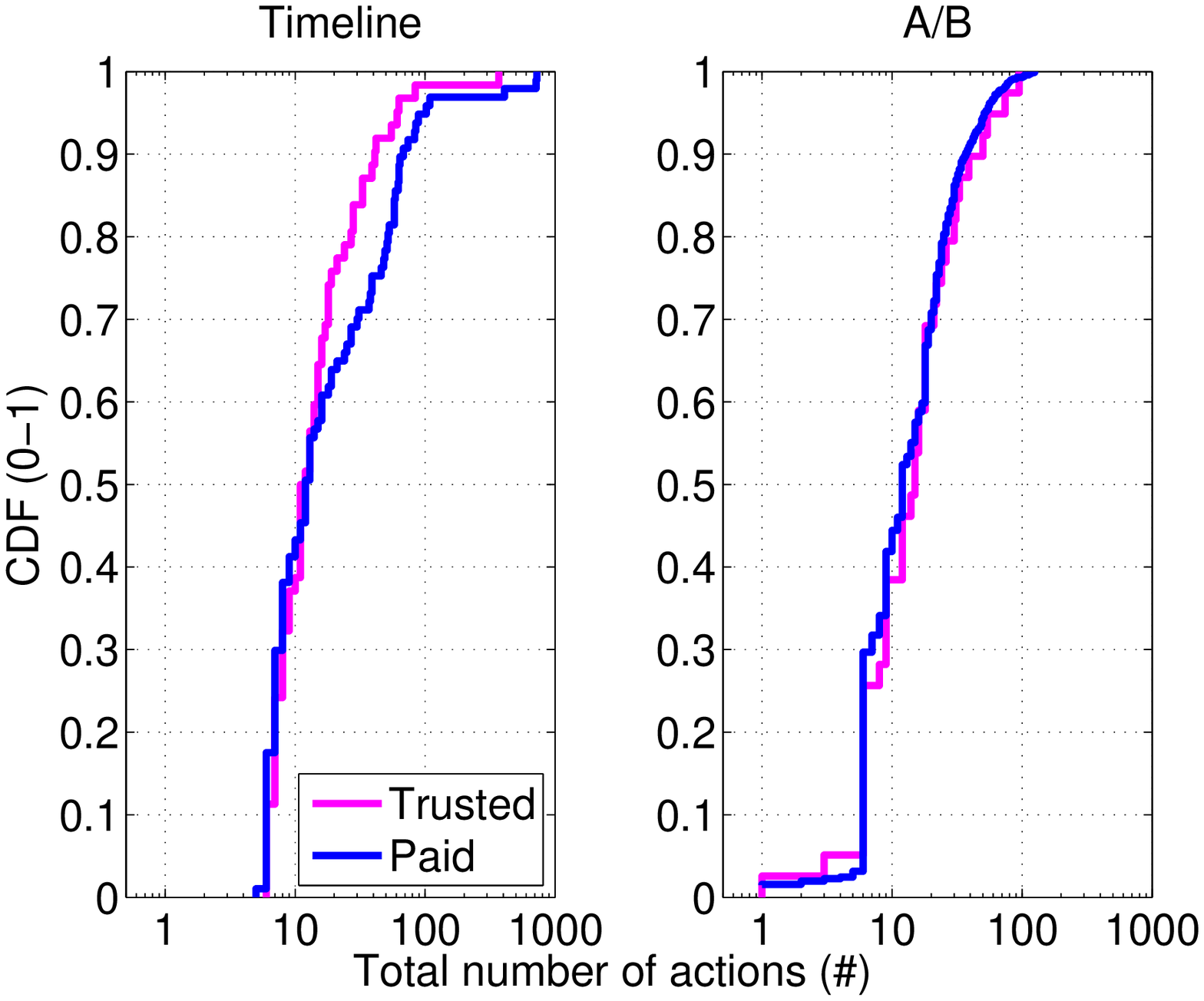, width=2.1in}
	\label{fig:validation-b}} 	
	\subfigure[Percentage of correct responses to control questions.]{\psfig{figure=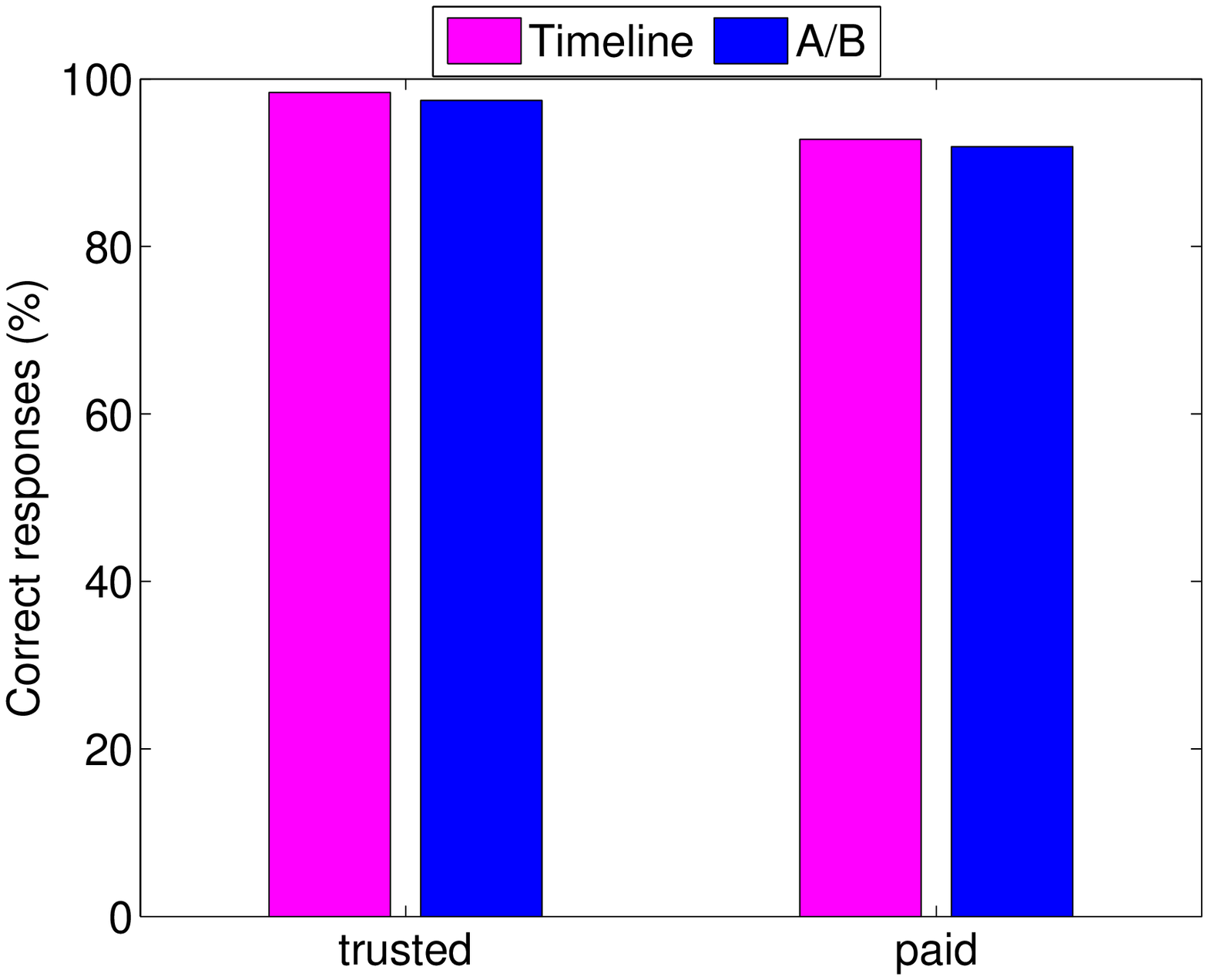, width=2.1in}
	\label{fig:validation-c}} 			
\caption{Participant behavior. \textnormal{Comparing the behavior of paid and trusted participants.}} \label{fig:validation} 
\end{figure*}

\subsection{Evaluating Filtering Techniques}
\label{sec:validation:results}
\mvnote{check if paid workers watch video for a longer time}

Based on the high-level techniques presented in \S\ref{sec:design-crowdsourcing}, our goal is to design a strategy for filtering out low-quality responses. We validate each technique by comparing paid responses after filtering to the trusted baseline.

\para{Engagement}~~ Figure~\ref{fig:validation-a} shows the Cumulative Distribution Function (CDF) of the time (minutes) each participant spent on \eo broken down both by participant type (paid and trusted) and experiment type (\timeline and A/B). The time spent on site is computed as the sum of the time spent evaluating each video, which in turn is computed as the difference between the time a response was submitted and the time the page loaded.

\begin{figure}[t] 
\centering 
	\psfig{figure=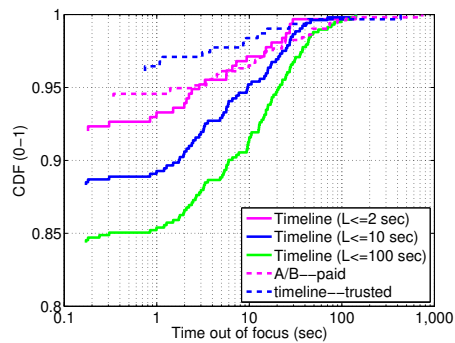, width=2.5in}	
	\caption{Out of focus time. \textnormal{$L =$ video load time.}}
	\label{fig:out-of-focus}
\end{figure}

Figure~\ref{fig:validation-a} shows that, for both experiment types, the CDFs for paid and trusted participants are quite similar,
though paid participants tend to take slightly longer than trusted participants (\eg the median grows from 2.5 to 3 minutes for the \timeline experiment). 
This result is counter-intuitive, as one would expect paid participants to skim through the videos faster in an attempt to increase their earnings by completing more tasks. Further analysis indicates that this is the result of two things: more ``out of focus'' time (Figure~\ref{fig:out-of-focus}) where \eo is in the background and longer video transfer times. The figure also suggests that the \timeline test requires more effort than the \AB test; it takes 3x longer on average. This is primarily because 1)~the \timeline experiment requires more interaction with the video (Figure~\ref{fig:validation-b}) to complete the task and 2)~the \timeline experiment requires the video to be fully loaded before the participant can begin the task (\S\ref{sec:design-experiments}).

Taking a closer look, Figure~\ref{fig:out-of-focus} shows the CDF of the time a participant did not have the \eo browser tab in the foreground. We do not plot results for trusted participants in the \AB experiments since only one participant switched tabs during that test. For the \timeline experiment (paid participants) we also differentiate by how long the video took to load ($L$). Overall, Figure~\ref{fig:out-of-focus} shows that participants tend to get more distracted the longer the video takes to load---we see about 10\% more distracted users when the video takes up to 100~seconds to load compared to less than 2~seconds. \AB campaign participants are, overall, as distracted as \timeline campaign participants for whom the video took less than 2 seconds to play. Since in the \AB campaign participants can click play right away and let the video load in the background, this further indicates that long video pre-loading time in the \timeline campaign is responsible for the (understandable) engagement drop. In addition, the higher complexity of the \timeline campaign might also have a role, as suggested by 4\% of the trusted participants also getting distracted for several seconds.

Next, we investigate how participants interact with each video; Figure~\ref{fig:validation-b} shows the CDF of the total number of actions (play, pause, and seek), again broken down by participant type and experiment type. The figure shows, overall, very similar CDFs, which indicates that paid and trusted participants are more similar in terms of actions performed than time spent on site. This further confirms the impact of external factors, such as long video load times, on the slower responses from paid participants. This result also shows \eo's ability to equalize participant feedback despite heterogeneous network conditions, which would be impossible if participants were asked to, for example, navigate to a given website and provide feedback. \mvnote{investigate and discuss crazy number of operations by paid participants.}

The tail of Figure~\ref{fig:validation-b} shows two users performing 714 and 724 actions (seek); this is twice as many actions as the most active trusted participant (369 seek actions). Nevertheless, these participants finished in 2 and 4 minutes, respectively, meaning that despite the large number of actions, they rank in the fastest 15\% and fastest 70\% of participants overall. Such frenetic behavior looks extremely suspicious, but, from our data, we could not figure out what causes it. We doubt it it is a realistic human interaction, and we conjecture a browser extension might have been used.

\begin{figure*}[bth] 
\centering 
	\subfigure[Sample \uplt CDFs]{\psfig{figure=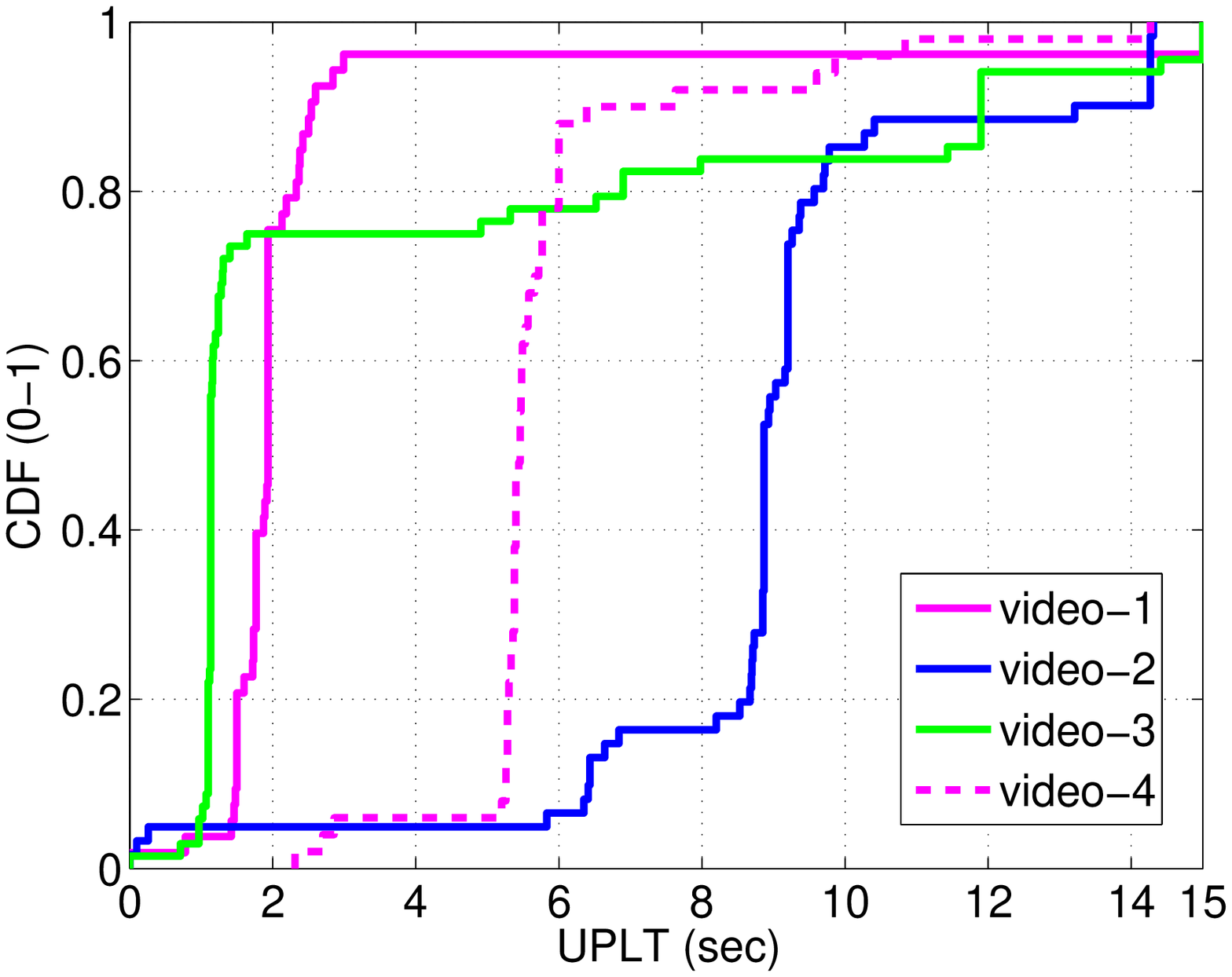,width=2.1in} 
	\label{fig:uplt-sample}}
	\subfigure[\Timeline experiments]{\psfig{figure=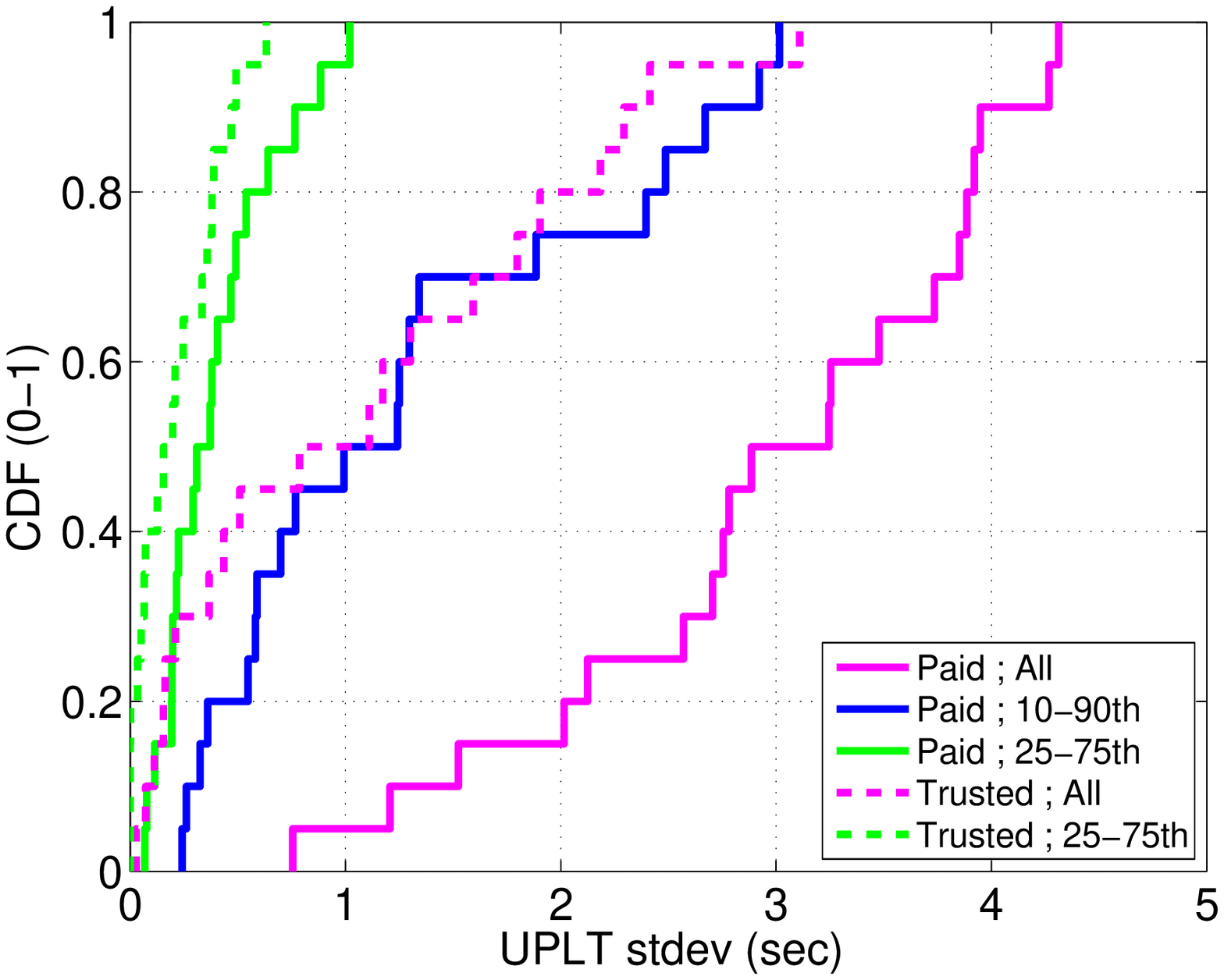, width=2.1in}
	\label{fig:wisdowm-crowd-timeline}} 	
	\subfigure[\AB experiments]{\psfig{figure=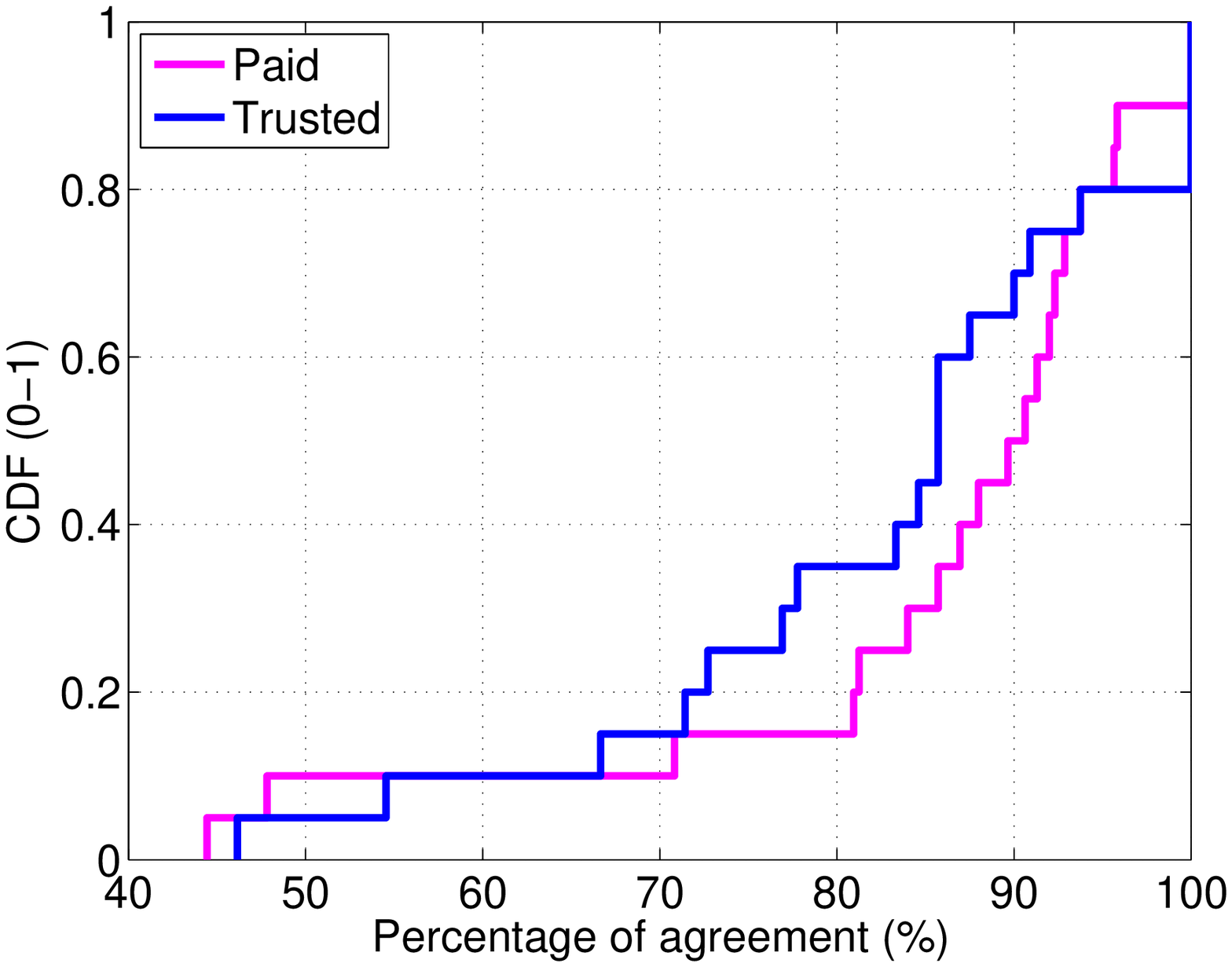, width=2.1in}
	\label{fig:wisdowm-crowd-ab}} 			
\caption{Wisdom of the crowd. \textnormal{With appropriate filtering, the responses of paid participants demonstrate a consistent majority opinion; furthermore, paid responses are in line with those of trusted participants.}} \label{fig:wisdom} 

\vspace{0.15cm}

\end{figure*}

\para{Soft Rules}~~Though \eo does not enforce it, we set a soft rule that participants should watch each video before submitting a response (\S\ref{sec:design}). Although hard to see due to log scale, Figure \ref{fig:validation-b} shows that trusted participants always interact with each video, seeking (\timeline) or hitting play (\AB) at least once. By contrast, 1--2\% of the paid participants speed through the test without interacting with the video at all. 

\para{Control}~~Figure \ref{fig:validation-c} shows the percentage of correct responses to control questions.
Overall, we notice no major difference between the two control types (\timeline versus \AB) for both paid and trusted participants, suggesting both control types were well designed. It is important to note, however, that paid participants tend to fail control questions at a higher rate (5\%), suggesting that there are some random clickers. Nevertheless, we found one distracted participant per campaign in the trusted user-set as well.

\para{Wisdom of the Crowd}~~ Finally, we investigate whether we can use consensus among paid participants as a proxy for ground truth in the absence of a trusted baseline, starting with the \timeline experiment.
To build intuition, Figure~\ref{fig:uplt-sample} shows the CDF of paid participants' \uplt for four representative websites. 
The wisdom of the crowd effect is clearly visible: the majority of the responses are concentrated around one (or a few) \uplt values, \eg 6.2 and 9.5 seconds for video-2. We further discuss sites with multiple modes (like the one shown in Figure~\ref{fig:viz-tool}) in \S\ref{sec:discussion}. 
The figure also shows fairly long heads/tails, indicating that, while most participants agree, some strongly disagree with the rest of the crowd. One possible explanation for this is that these participants simply scroll to the beginning or end of the video in an attempt to finish the test and get paid as soon as possible.

We generalize this observation to the entire \timeline campaign by using the standard deviation of the \uplt for each website as a measure of agreement among parti\-cipants---the tighter the distribution, the more in agreement the responses are.
By removing outliers at the ends of each distribution (\eg keeping only responses between the 10th and 90th percentiles of each site's \uplt distribution), we arrive at a set of high-quality responses in relatively tight agreement.
Figure~\ref{fig:wisdowm-crowd-timeline} shows CDFs of standard deviations across all videos in the validation data set.
First, without any filtering, the ``Paid All'' and ``Trusted All'' show wide standard deviations, which comes as no surprise based on the tails in Figure~\ref{fig:uplt-sample}). 
Second, the gap between the ``Paid All'' and ``Trusted All'' curves demonstrates the overall lower quality of the paid participants. 
Third, as we apply filtering, we see that the standard deviation of \uplt drops quickly, indicating most participants are in agreement.
Finally, when restricted to responses between the 25th and 75th percentiles, results for paid and trusted participants are in line, which confirms that the wisdom of the paid crowd is a reasonable pseudo-ground truth in the absence of a trusted baseline.

Next, we extend this analysis to the \AB experiment. In this case, the analysis is simpler as the user input is restricted to three discrete values---left was faster, right was faster, or there was no difference---rather than the continuous set of values for the \timeline experiment. Accordingly, Figure~\ref{fig:wisdowm-crowd-ab} shows the CDF of participant ``agreement'' for each video for paid and trusted participants. We define agreement as the fraction of responses matching the most popular answer, independent of what that answer is. For example, an agreement of 80\% indicates that 80\% of the votes went to one choice and the remaining 20\% were split, somehow, between the other two.

Figure \ref{fig:wisdowm-crowd-ab} shows no dramatic difference between trusted and paid participants in term of overall agreement. Likely, this is due to the ease of the \AB task compared to the \timeline task, as previously discussed. The figure also shows a high level of agreement, \eg more than 85\% of participants converge to the same response for 60\% of the videos and 10--20\% of the videos have a 100\% agreement. In addition, we never saw a completely split response (33\% agreement)---the minimum level of agreement in the figure is 45\%.  Paid participants tend to agree more than trusted participants. Coupled with a lower rate of ``no difference'' responses, this might indicate that trusted participants are more cautious about their choices.

 \begin{figure*}[htb] 
\centering 
	\subfigure[Comparison of submitted \uplt with slider selection and frame-helper suggestion.]{\psfig{figure=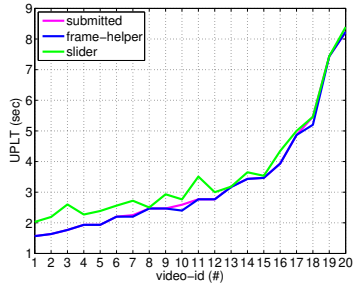, width=2.1in}
	\label{fig:plt-eval-a}} 
	~~~~
	\subfigure[Scatter-plot of \uplt and PLT metrics.]{\psfig{figure=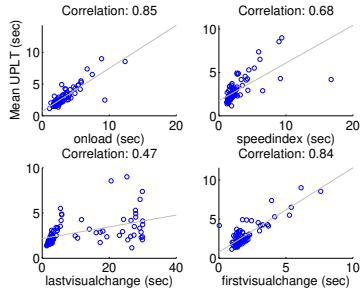,width=2.1in} 
	\label{fig:plt-eval-b}}	
	~~~~
	\subfigure[CDF of difference between \uplt and PLT metrics.]{\psfig{figure=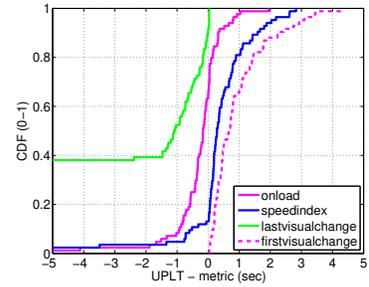, width=2.1in}
	\label{fig:plt-eval-c}} 				
\caption{Timeline Results.} 
\label{fig:plt-eval} 
\end{figure*}

\newpage
\subsection{Final Filtering Strategy}
\label{sec:cleaning}
\label{sec:sec:cleaning}
Based on the analysis above, we arrive at the following methodology for filtering low-quality responses.  Table~\ref{tab:summary} (last three columns) summarizes the number of participants dropped per campaign according to the methods below.

\para{Engagement}~~ The previous analysis suggests that the time spent on site (Figure~\ref{fig:validation-a}) is not a reliable way to filter unengaged participants as it can be impacted by external factors like network connectivity. 
Instead, we use the following two engagement metrics.
First, a large number of video interactions (Figure~\ref{fig:validation-b}) is a good indicator of suspicious activity, so we introduce a simple rule dropping paid participants with 50\% more video interactions than the most active trusted participant (369 seeks). 
This filtering is very minor; it applied to only two paid participants in the validation data-set and three paid participants\footnote{These paid participants managed to perform between 912 and 1931 seek actions in 5 minutes.} in the final data-set (\S\ref{sec:plt-results-data}).
Second, we apply a more aggressive filter based on the amount of time a participant is distracted, with the caveat that longer video load times may understandably cause participants to briefly switch to other tabs. 
Accordingly, we filter participants who switch away from the \eo tab for more than 10 seconds so long as the video was delivered within those 10 seconds. 
Table~\ref{tab:summary} shows that this removes 10--15\% of the paid participants. 

\para{Soft Rules}~~ We discard responses from participants who skipped (\ie did not play or scrub) even just one video. This removes 2--5\% of both trusted and paid participants (apart from the trusted \timeline campaign, where no participant broke this rule).

\para{Control}~~ We discard participants who failed any control question. This amounts to 2--8\% of participants, with no consistent trend between campaign types (more \timeline participants fail the control in the validation set, whereas more \AB participants fail in the final data set). Only one trusted participant failed a control question in each campaign, which gives us confidence our control mechanisms do not introduce false positives.

\para{Wisdom of the Crowd}~~ For \timeline campaigns, we only keep responses between the 25th and 75th percentiles for each video.

\newcommand{\minviews}{{3}\xspace}
\newcommand{\ab}{{AdBlock}\xspace}
\newcommand{\gh}{{Ghostery}\xspace}
\newcommand{\ub}{{uBlock}\xspace}

\ourcomment{
	\begin{table*}[b]
	\centering
	\caption{Workers Feedback}
	\label{tab:workers}
	\begin{tabular}{|l|l|l|l|l|l|}
	\hline
	               & \textbf{Overall} & \textbf{Instructions Clear} & \textbf{Test Questions Fair} & \textbf{Ease of Job} & \textbf{Pay} \\ \hline
	\textbf{PROTO} & 4.6/5            & 4.6/5                       & 4.6/5                        & 4.6/5                & 4.6/5        \\ \hline
	\textbf{H1-H2} & 4.6/5            & 4.6/5                       & 4.6/5                        & 4.6/5                & 4.6/5        \\ \hline
	\textbf{ADS}   & 4.6/5            & 4.6/5                       & 4.6/5                        & 4.6/5                & 4.6/5        \\ \hline
	\end{tabular}
	\end{table*}
}

\section{Page Load Time Results}
\label{sec:plt-results}

This section summarizes our findings from using \eo to study page load time (PLT), an important aspect of QoE. Specifically, we answer three questions: 1) to what extent do existing PLT metrics represent user perception, 2) do users perceive a speed difference between HTTP/1.1 and HTTP/2, and 3) what impact do ad blockers have on PLT?

\subsection{Data Set}
\label{sec:plt-results-data}
To answer these questions, we ran three campaigns, summarized in the \emph{Final} section in Table~\ref{tab:summary}. Each campaign consisted of 100 videos and targeted 1,000 paid participants (again, recruited from \cf's ``most trustworthy'' pool). Each participant watched six videos, meaning we served 6,000 videos in total per campaign and each video has, on average, responses from 60 different participants. 

Overall, it took about 1.5 days per campaign to recruit 1,000 participants at a cost of \$120. Across all three campaings, 70\% of the participants are male and 30\% are female. Participants are located in 76 distinct countries, with Venezuela being the most popular.

\subsection{User Perception and PLT}

We compare \uplt against four existing, automatically computable metrics:

\para{\ol}~refers to the time it takes for the JavaScript ``onLoad'' event to fire.
This event fires once the page's \emph{embedded} resources have been downloaded, but not necessarily before, \eg objects loaded via scripts are retrieved.
\ol is the simplest and most used metric in practice.

\para{\si}~``is the average time at which visible parts of the page are displayed'' (the average is over all above-the-fold pixels).
Imagine a curve plotting the percentage of pixels that are ``visually complete'' (\ie match their final state) over time as the page loads; \si is the area \emph{above} this curve (smaller values better---the load feels faster the more ``up and to the left'' the curve is).

\para{\fvc/\lvc}~are the times at which the first pixels are drawn and the last pixels stop changing on the user's screen. \\

Figure \ref{fig:plt-eval} summarizes the results from the \timeline campaign.  For each video, we compute the \uplt as the mean of the filtered participant responses.
To start, we investigate the impact of the frame selection helper (\S\ref{sec:design-experiments}). Figure \ref{fig:plt-eval-a} compares participants' final \uplt choices (``submitted'') with their original choices (``slider'') and the one proposed by the frame selection dialog (``frame-helper''). For visibility, we only show results for the 20 videos used for validation in Section \ref{sec:validation}. For most videos, the submitted \uplt matches the value proposed by the frame selection helper, suggesting that most participants agree with \eo's suggestions for fine-tuning their responses. On average, the submitted \uplt differs from the original value selected with the slider by 300~ms and up to a maximum of 1.6~s. In the remainder of the PLT analysis, we only consider the submitted \uplt and extend to the full video pool. 
\mvnote{manually investigate videos where it was not used}

\begin{figure*}[htb] 
\centering 
	\subfigure[Agreement as a function of PLT $\Delta$.]{\psfig{figure=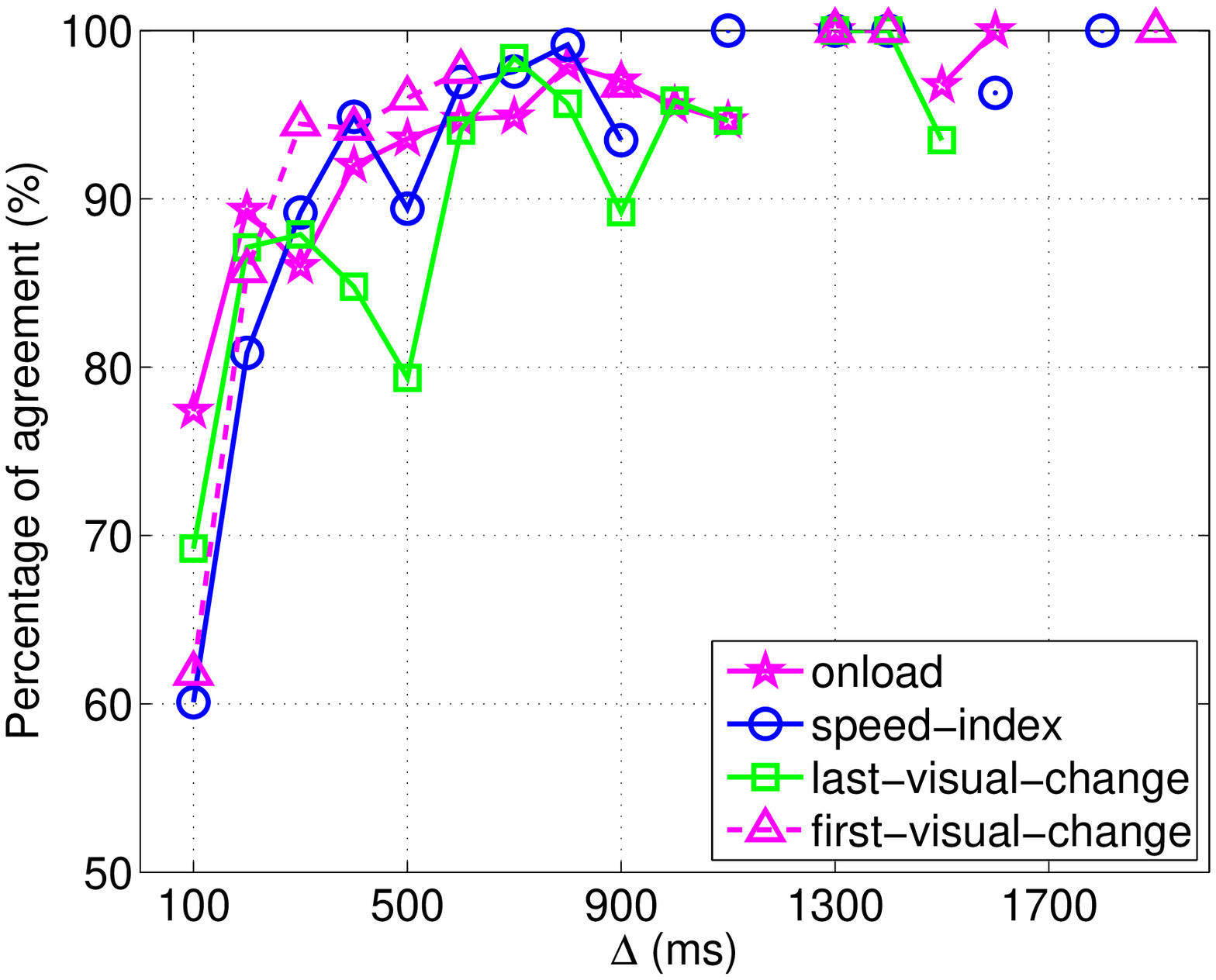, width=2.1in}
	\label{fig:proto-eval-a}} 	
	\subfigure[\h vs \hh.]{\psfig{figure=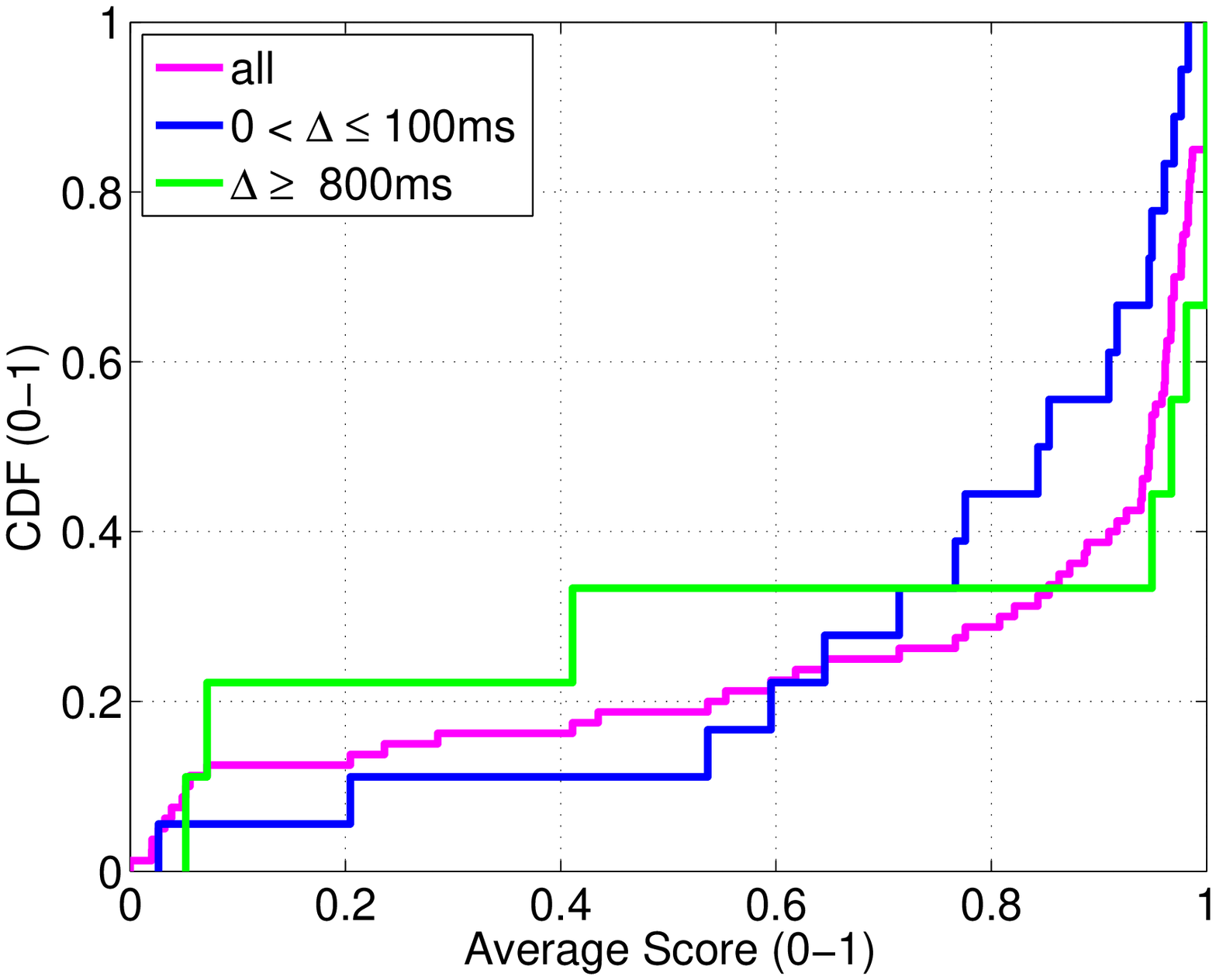,width=2.1in} 
	\label{fig:proto-eval-b}}
	\subfigure[Comparison of ad blockers.]{\psfig{figure=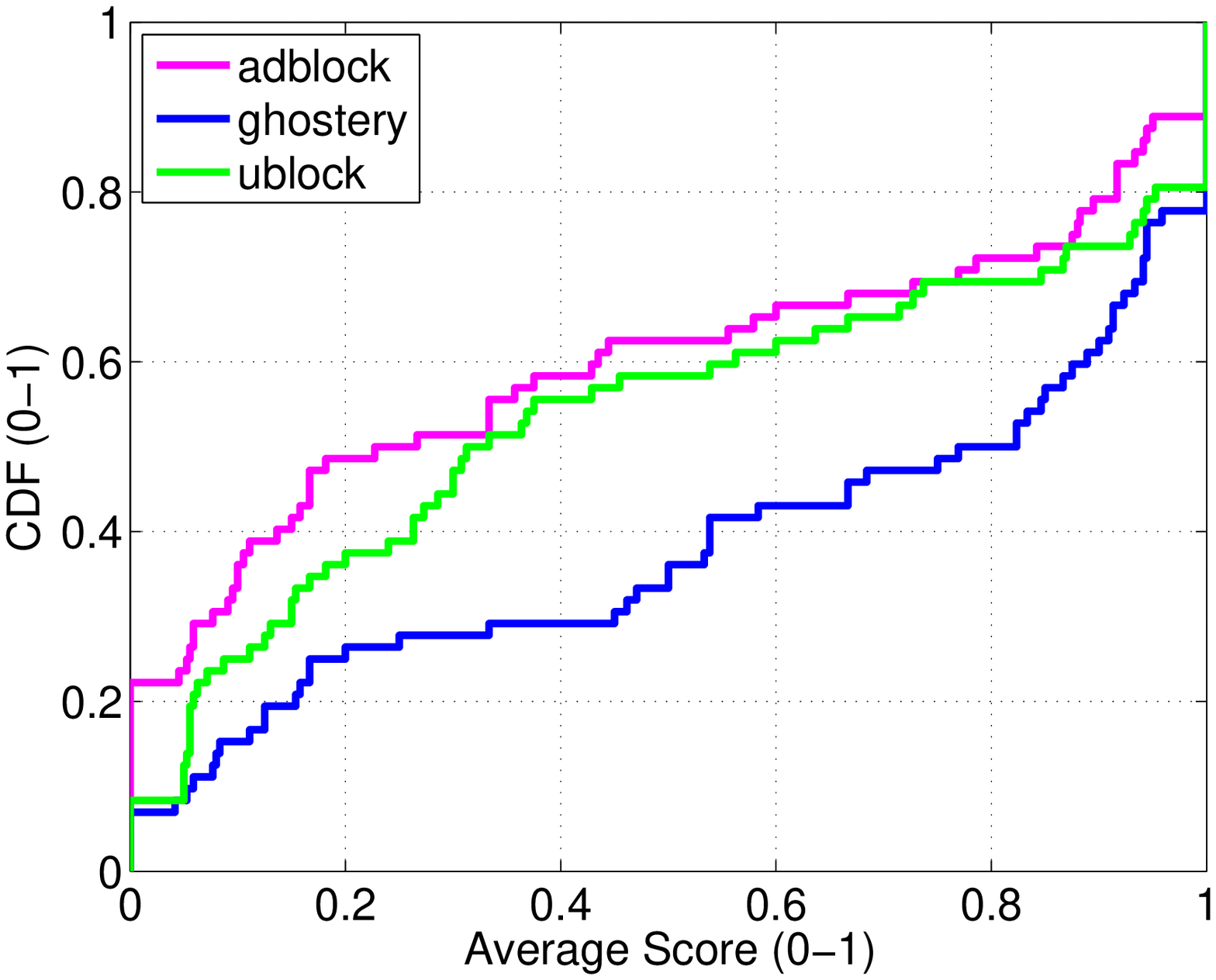, width=2.1in}
	\label{fig:proto-eval-c}} 
\caption{\AB Results.} 
\label{fig:proto-eval} 
\vspace{-0.15in}
\end{figure*}

\vspace{0.2cm}
\paraittight{\textbf{Are existing metrics related to UserPerceivedPLT at all?}}~~\\
Figure~\ref{fig:plt-eval-b} shows how \uplt correlates with each of the four PLT metrics. \ol and \fvc highly correlate with \uplt (0.84--0.85), while \si, despite its higher complexity, has a lower correlation with \uplt (0.68). \lvc provides almost no indication of when participants consider a page to be loaded (0.47). Though a non negligible number of responses (7\%) exactly match \lvc, most participants indicate that the pages are ``ready enough'' to be used way earlier.

\newpage
\paraittight{\textbf{Are existing metrics able to predict the \emph{value} of UserPerceivedPLT?}}~~
While correlation gives a rough picture of how each metric relates to \uplt, it is also important to show how close the actual values are. Figure~\ref{fig:plt-eval-c} shows a CDF of the difference between \uplt and each PLT metric, \ie a negative number indicates that participants selected a time \emph{before} that metric indicated the page had loaded. Overall, \ol was within 100~ms of the mean \uplt for 30\% of the sites compared to just 7\% for \si. Also note that 60\% of the \uplt values are smaller than \ol---that is, \ol tends to over-estimate. Unsurprisingly, \fvc and \lvc under- and over-estimate, respectively.

\paraittight{\textbf{Can existing metrics at least tell us ``which is faster''?}}~~
Finally, we quantify how good the existing PLT metrics are at identifying a difference in load time between two page loads using data from the \h vs \hh \AB campaign.
Intuitively, choosing which site in an \AB test is faster becomes easier as the absolute difference between the ``true'' load times for $A$ and $B$ ($\Delta$) increases. 
Therefore, we can use the level of participant agreement as a proxy for $\Delta$---the more the participants agree, the larger $\Delta$ likely was.

Based on this reasoning, 
1) if a PLT metric is good, we expect to see agreement increase monotonically as the measured $\Delta$ for that metric increases, and 
2) the higher the agreement for a particular $\Delta$-value for a particular metric, the more confidence we have that when we measure a $\Delta$ of that size in the future, it represents a meaningful difference in \uplt.
Figure~\ref{fig:proto-eval-a} shows the median agreement (as previously defined in \S\ref{sec:validation:results} Figure~\ref{fig:validation-c}) among participants for \h/\hh video pairs as a function of each metric's $\Delta$.
Overall, Figure~\ref{fig:proto-eval-a} shows that as $\Delta$ increases, participants tend to agree more, 
which matches our intuition.
While there is no clear winner among the PLT metrics, the figure shows that \ol better captures small loading time differences ($\Delta \leq 200$~ms) whereas \si and \fvc do a better job in the medium range  ($200 < \Delta \leq 800$~ms). 
The figure also shows that \lvc and \si do not exhibit a monotonic increase in agreement as $\Delta$ grows, meaning that small variations in \lvc and \si measurements are less significant.

\mvnote{here add a connection with discussion}

\ourcomment{
	\begin{figure*}[htb] 
	\centering 
		\subfigure[PLT metrics comparison.]{\psfig{figure=\folder/index-analysis-PROTO.eps, width=2.1in}
		\label{fig:proto-eval-a}} 			
		\subfigure[\h vs \hh.]{\psfig{figure=\folder/h1-h2_speed-index.eps,width=2.1in} 
		\label{fig:proto-eval-b}}
		\subfigure[\ab vs \gh vs \ub]{\psfig{figure=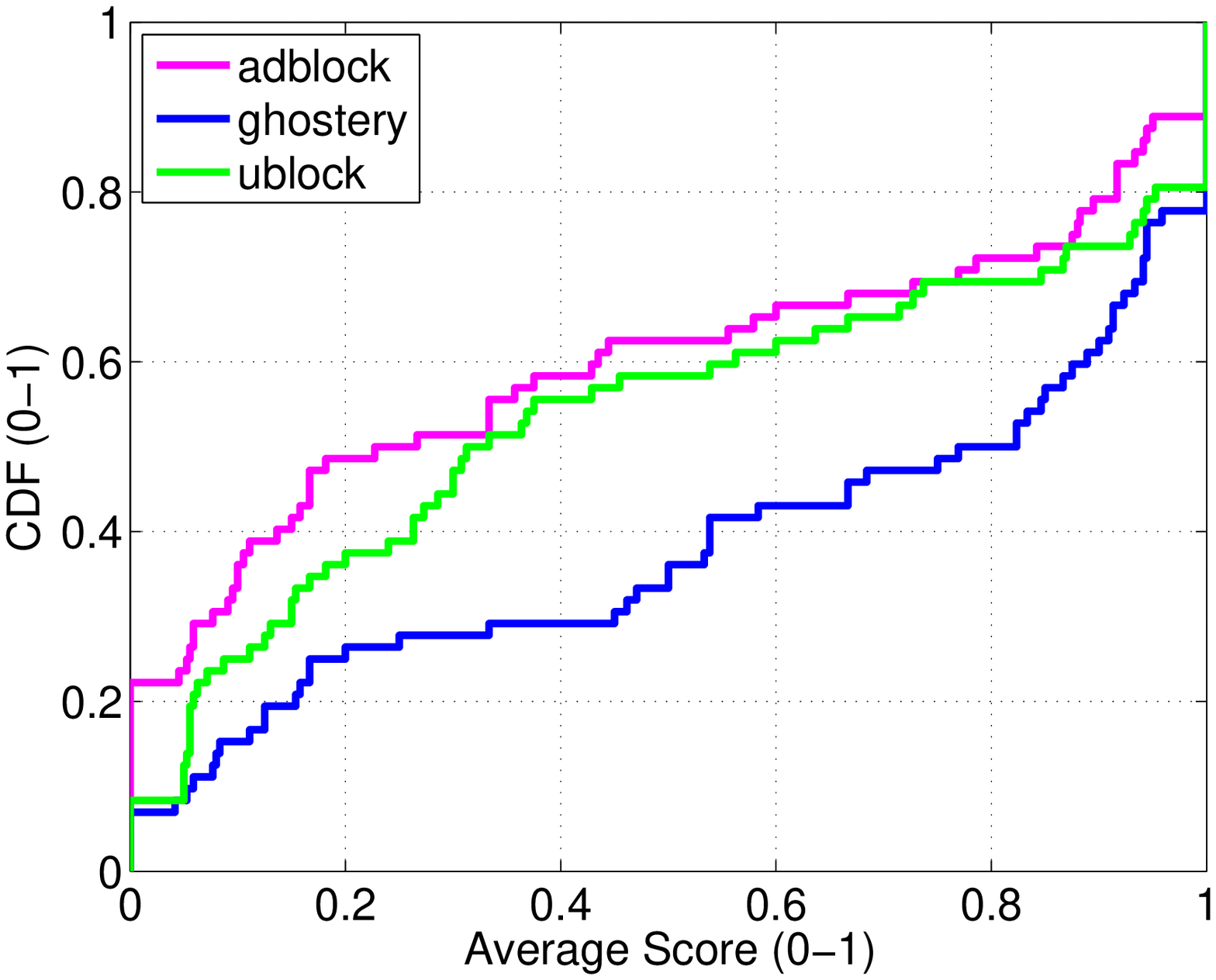, width=2.1in}
		\label{fig:proto-eval-c}} 
	\caption{\proto Results} 
	\label{fig:proto-eval} 
	\end{figure*}
}

\subsection{\h vs \hh}
\paraittight{\textbf{Do users perceive a speed difference between \h and \hh?}}~~
In this section, we examine the responses from our \h vs \hh \AB campaign. Figure \ref{fig:proto-eval-b} shows the CDF of the average ``score'' per website; 0 means the \h version was faster, 0.5 is a ``split'' decision, and 1 means the \hh version was faster. We plot scores for 1) all websites, 2) websites with similar \h and \hh PLTs ($\Delta \leq 100$~ms),  and 3) websites that loaded much faster over one protocol than the other ($\Delta \geq 800$~ms). To build these subsets, we compute PLT using \si. 

Figure \ref{fig:proto-eval-b} shows that 70\% of the websites have an average score of 0.8 or higher; this means that 70 out of 100 websites ``feel'' faster using \hh than \h. Conversely, 12\% of the websites have an average score of 0.2 or lower and thus feel faster using \h. The remaining 18\% of websites create some disagreement. Note that the score here does not take into account the ``No Difference'' responses. These websites with scores in the 0.2--0.8 range also have twice as many No Difference responses compared to the other websites. This further indicates that participants are just not sure which version was actually faster.

Next, we focus on the subset of websites with similar PLTs ($\Delta \leq 100$~ms). The figure shows that participant indecision grows, with more scores in the 0.2--0.8 range. This is to be expected based on the results from Figure~\ref{fig:proto-eval-a}. On the other hand, when $\Delta \geq 800$~ms, participants mostly agree on which version was faster. This result indicates that, while aiming at reducing loading time of a webpage is overall beneficial, many users are not able to appreciate the difference when only few hundred milliseconds are saved.

\subsection{Ad Blocker Comparison}

\paraittight{\textbf{How do popular ad blockers impact PLT?}}~~
We compare three popular ad blockers, \ab, \gh, and \ub. Figure \ref{fig:proto-eval-c} shows the CDF of the average ``score'' obtained by each website where 0 means the original version with ads was faster and 1 means the ad-blocked version was faster.  

The figure shows that 30--40\% of the websites have scores in the 0.2--0.8 range, \ie participants did not agree on which version was faster. This is about 15\% more compared to when participants were asked to evaluate \h versus \hh. Based on feedback collected on \eo, we believe this is due to the fact that the two versions of the websites are now not perfectly equal, which makes deciding which was faster harder. Nevertheless, the figure shows \gh is a clear favorite; for example, for 50\% of sites, participants strongly agreed ($\geq 0.8$) that the \gh version of the page was faster compared to just 25\% for \ab and \ub.

\ourcomment{
	\begin{figure*}[htb] 
	\centering 
		\subfigure[\h vs \hh.]{\psfig{figure=\folder/speed-index.eps,width=2.1in} 
		\label{fig:eval_a}}
		\subfigure[PLT metrics comparison.]{\psfig{figure=\folder/index-analysis-PROTO.eps, width=2.1in}
		\label{fig:eval_b}} 	
		\subfigure[...]{\psfig{figure=\folder/index-analysis-ADS.eps, width=2.1in}
		\label{fig:eval_c}} 	
	\caption{\proto Results} 
	\label{fig:evaluation} 
	\end{figure*}
}

\section{Discussion}
\label{sec:discussion}

\para{What Does ``Ready'' Mean?}~~In our \timeline tests, we ask participants
to ``drag the slider to the point where you consider the site `ready to use.'''
We intentionally left the wording open to individual interpretations of
``ready,'' since \emph{what humans consider ``ready'' is exactly what we're
trying to learn}. If, for example, we instructed participants to ``pick the
point where content stops changing,'' we would simply be asking them to
reproduce \lvc.

\begin{figure}[t]
\centering
\includegraphics[width=\columnwidth]{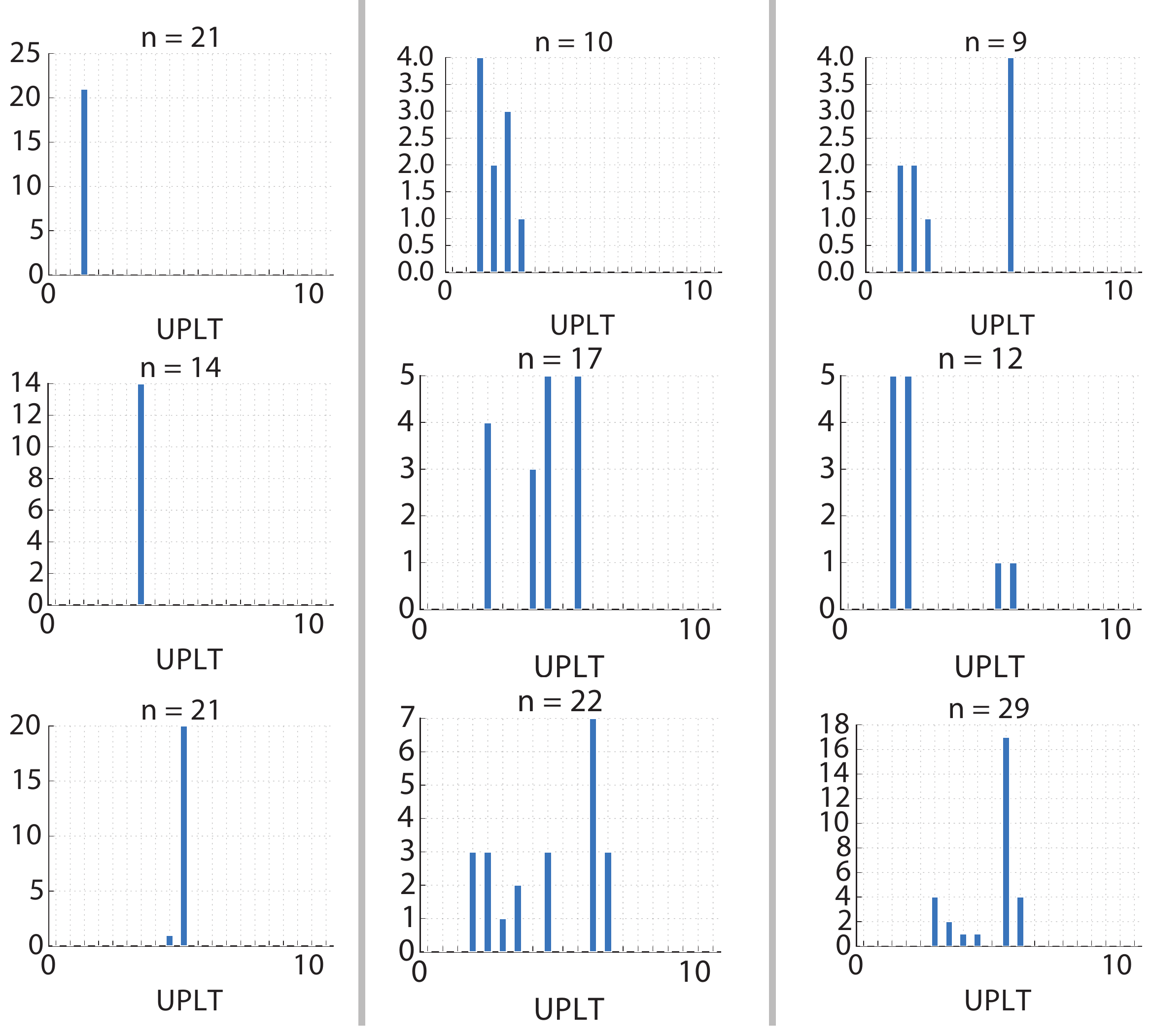}
\caption{\uplt Distributions: \textnormal{Sites typically had one of three
types of distributions: single mode and tight (left column), single mode and
spread out (center column), or multi-modal (right column).}}
\label{fig:distributions}
\end{figure}

To see if participants have consistent definitions of ``ready,'' we look at the
\uplt distributions for different sites.  Three rough patterns emerge
(Figure~\ref{fig:distributions}).
1)~Some sites exhibit a single, clear peak in \uplt choices.
After manually inspecting the associated videos, these sites tend to be characterized by ``fast'' loads (in the sense that the span of time between \fvc and \lvc is very short). 
These cases are relatively cut-and-dry;
not much is open to interpretation, and participants are pretty consistent with the times they choose.
2)~Some sites have a much wider distribution.
These sites tend to have a much
longer gap between \fvc and \lvc, giving participants more freedom to choose
different ``ready'' times.
3)~Some sites exhibit \emph{multiple} peaks.
In some cases, this appears to be due to auxiliary content, like social media widgets and
ads---some participants consider the page ready after the main content has
loaded, while others wait for the auxiliary content to load.

Comments we collected from trusted participants help shed light on this.  The first two
comments confirm that auxiliary content does impact some participants'
responses:

\begin{quote} \small
\textit{
``Sometimes my choice may not be ... due to the [order] in which the content
loaded. I selected the one where the main content loaded first, not necessarily
which one finished loading everything [first].''
}
\end{quote}

\begin{quote} \small
\textit{
``Ads/like buttons/etc. usually load last; it's a little unclear whether
`ready to use' should include those or not. I start using sites before all of
the ads loaded, but when I do that, I know that the page isn't totally done
loading---I just don't care.''
}
\end{quote}

In conjunction with Figure~\ref{fig:distributions} these comments are particularly interesting in light of new features in HTTP/2 like push and priority that could be exploited to ensure the optimal delivery order for \emph{specific} users (similar to Klotski~\cite{klotski}).

The following two comments echo the first two, but also point out some limitations
of our methodology. First, some participants' responses are affected by their
familiarity with a site.
They are easily able to identify when the important content is loaded if the site is familiar (\ie they will think the page is loaded ``faster''), whereas with an unfamiliar site they might wait until every piece of content loads.
Second, some participants look for more than just
visual cues, \eg when an element on the page becomes responsive to user
input. \eo's page load videos fail to capture these cues.

\begin{quote} \small
\textit{
``Sometimes how early I feel the site is ready to use depends on what I want to
use the site for. For example, if I am looking for articles, I don't care if
the ads have loaded. So usually I consider the webpage ready when the part I
want has loaded. However, there are other cases, like I don't know what is on
the site, when I want to wait for everything to load. There is also the case
where the links move when other parts load and for those cases, I commonly have
to wait for the whole page to load because it's too easy to click where the
link used to be and click something else instead.''
}
\end{quote}

\begin{quote} \small
\textit{
``It is difficult to say when a site becomes `usable' without knowing whether the search bar is responsive. For many sites I consider it usable the instant the search bar becomes responsive.''
}
\end{quote}

\para{Extending Eyeorg}~~In this paper, we only scratched the surface of what can be done with \eo.
Many \emph{existing} features of \eo that enable rich experimentation, \eg (mobile) device and network emulation, have not been discussed at all.
Similarly, although we have provided an extensive analysis into understanding how humans perceive PLT, we are fully aware that \emph{many} open questions remain.
We believe that the community at large can leverage \eo as a platform to enable high quality, novel research that answers these and other questions.
\eo can be used to measure the impact of a variety of changes to the web; \eg TCP vs. QUIC, TLS 1.2 vs TLS 1.3, \hh push/priority strategies, web design techniques like domain sharding or image spriting, browser plugins, or even in-network services like Google's Flywheel compression proxy~\cite{flywheel}.

\para{Data Collection and Privacy}~~
\eo collects data from human participants, so we have to take care we collect and use this data responsibly. (Note that paid participants may also provide data directly to the crowdsourcing services \eo uses; this data is not shared with us and is subject to those services' privacy policies.)

We collect three types of data. First, we ask participants to provide standard demographic information like gender, age, country, and self-assessed technical ability. We collect this information at a coarse enough granularity there is no danger of identifying individual people.
Next, we collect activity data like how long a participant spends reading instructions, when they play/pause a video,  and whether they switch away from the \eo tab during an experiment.
(Note that we only collect information about the \eo tab; if they switch away, we have no information about what they do.)
We use this data to filter out uncommitted participants.
We do \emph{not} use this data to discriminate (\eg refuse payment) against participants.Finally, we collect participant responses to our QoE questions (\eg ``when is this page ready to use'' or ``which video is faster''). Clearly this data is needed, as it is what \eo seeks to measure to begin with.

\section{Conclusion}
\label{sec:conclusion}

This work presented \eo, a platform for crowdsourcing web quality of experience measurements. 
\eo collects \emph{quantitative} QoE measurements from real users \emph{at scale}.
\eo relies on \emph{videos} to ensure each participant sees a consistent view of the page loading; our video collection tool control parameters like device type, protocol, network conditions, and more.
We validate the responses collected from crowdsourced workers by comparing them to responses from a controlled set of 100 trusted users.
Next, we conduct three measurements campaigns involving 1,000 participants each; our campaigns  investigate the quality of existing PLT metrics, compare HTTP/1.1 vs HTTP/2 performance, and assess the impact of online advertisements on user experience. 
Going forward, we plan to extend \eo with a broader set of capabilities for conducting web QoE experiments.
In particular, we plan to make \eo a platform that any researcher can use to test web content or delivery optimizations easily, without worrying about the challenges of designing a user study from scratch.

\section*{Acknowledgements}
This project has received funding from the European Union's Horizon 2020 research and innovation programme under grant agreement No 688421. The opinions expressed and arguments employed reflect only the authors' view. The European Commission is not responsible for any use that may be made of that information.

\bibliographystyle{acm}

\begin{thebibliography}{10}

\bibitem{onload_mdn}
Globaleventhandlers.onload.
\newblock
  \url{https://developer.mozilla.org/en/docs/Web/API/GlobalEventHandlers/onload}.

\bibitem{plt_revenue}
{How One Second Could Cost Amazon \$1.6 Billion In Sales}.
\newblock
  \url{http://www.fastcompany.com/1825005/how-one-second-could-cost-amazon-16-billion-sales}.

\bibitem{h2_page}
{Is the Web HTTP/2 Yet?}
\newblock \url{http://isthewebhttp2yet.com}.

\bibitem{flywheel}
{\sc Agababov, V., Buettner, M., Chudnovsky, V., Cogan, M., Greenstein, B.,
  McDaniel, S., Piatek, M., Scott, C., Welsh, M., and Yin, B.}
\newblock Flywheel: Google{\textquoteright}s data compression proxy for the
  mobile web.
\newblock In {\em Proc. USENIX NSDI\/} (Oakland, CA, USA, May 2015).

\bibitem{arapakis2014searchlatency}
{\sc Arapakis, I., Bai, X., and Cambazoglu, B.~B.}
\newblock Impact of response latency on user behavior in web search.
\newblock In {\em Proc. ACM SIGIR\/} (Gold Coast, Queensland, AU, July 2014).

\bibitem{angeles2015physiologicallatency}
{\sc Barreda-\'{A}ngeles, M., Arapakis, I., Bai, X., Cambazoglu, B.~B., and
  Pereda-Ba\~{n}os, A.}
\newblock Unconscious physiological effects of search latency on users and
  their click behaviour.
\newblock In {\em Proc. ACM SIGIR\/} (Santiago, Chile, Aug. 2015).

\bibitem{blackburn2014stfunoob}
{\sc Blackburn, J., and Kwak, H.}
\newblock {{S}{T}{F}{U} {N}{O}{O}{B}! Predicting Crowdsourced Decisions on
  Toxic Behavior in Online Games}.
\newblock In {\em Proc. WWW\/} (Seoul, Korea, Apr. 2014).

\bibitem{dario2016webqoe}
{\sc Bocchi, E., De~Cicco, L., and Rossi, D.}
\newblock Measuring the quality of experience of web users.
\newblock In {\em In Proc. LANCOMM\/} (Florianopolis, Brazil, Aug. 2016).

\bibitem{bouch2000webqoe}
{\sc Bouch, A., Kuchinsky, A., and Bhatti, N.}
\newblock Quality is in the eye of the beholder: Meeting users' requirements
  for internet quality of service.
\newblock In {\em CHI\/} (The Hague, The Netherlands, Apr. 2000).

\bibitem{klotski}
{\sc Butkiewicz, M., Wang, D., Wu, Z., Madhyastha, H.~V., and Sekar, V.}
\newblock Klotski: Reprioritizing web content to improve user experience on
  mobile devices.
\newblock In {\em Proc. USENIX NSDI\/} (Oakland, CA, USA, May 2015), NSDI'15.

\bibitem{Egger2012ICC}
{\sc Egger, S., Reichl, P., Hobfeld, T., and Schatz, R.}
\newblock "time is bandwidth"? narrowing the gap between subjective time
  perception and quality of experience.
\newblock In {\em Proc. IEEE ICC\/} (Ottawa, Canada, June 2012).

\bibitem{eyeorg_url}
{\sc Eyeorg}.
\newblock A platform for crowdsourcing web quality of experience measurements.
\newblock \url{https://eyeorg.net}.

\bibitem{crowdflower.shits.on.mturk}
{\sc Harris, A.}
\newblock Dropping mechanical turk helps our customers get the best results,
  Jan. 2014.
\newblock \url{https://speedcurve.com}.

\bibitem{Hossfeld2014}
{\sc Hossfeld, T., Keimel, C., Hirth, M., Gardlo, B., Habigt, J., Diepold, K.,
  and Tran-Gia, P.}
\newblock Best practices for qoe crowdtesting: Qoe assessment with
  crowdsourcing.
\newblock {\em Trans. Multi. 16}, 2 (Feb. 2014).

\bibitem{repos}
{\sc Lidskog, T.}
\newblock browsertime.
\newblock \url{https://github.com/tobli/browsertime}.

\bibitem{speedcurve}
{\sc Limited, S.}
\newblock Speedcurve.
\newblock \url{https://speedcurve.com}.

\bibitem{polaris}
{\sc Netravali, R., Goyal, A., Mickens, J., and Balakrishnan, H.}
\newblock Polaris: Faster page loads using fine-grained dependency tracking.
\newblock In {\em Proc. USENIX NSDI\/} (Santa Clara, CA, USA, Mar. 2016).

\bibitem{mobilyzer}
{\sc Nikravesh, A., Yao, H., Xu, S., Choffnes, D., and Mao, Z.~M.}
\newblock Mobilyzer: An open platform for controllable mobile network
  measurements.
\newblock In {\em Proc. ACM MobiSys\/} (Florence, Italy, May 2015).

\bibitem{Rainer2015}
{\sc Rainer, B., Waltl, M., and Timmerer, C.}
\newblock {\em A Web based Subjective Evaluation Platform}.
\newblock Klagenfurt am Worthersee, Austria, July 2013.

\bibitem{Rzeszotarski2011}
{\sc Rzeszotarski, J.~M., and Kittur, A.}
\newblock Instrumenting the crowd: Using implicit behavioral measures to
  predict task performance.
\newblock In {\em UIST\/} (Santa Barbara, California, USA, May 2011).

\bibitem{measuring.performance}
{\sc Shull, M.}
\newblock Measuring performance, Apr. 2015.
\newblock \url{https://davidwalsh.name/measuring-performance}.

\bibitem{bookofspeed}
{\sc Stefanov, S.}
\newblock Book of speed.
\newblock \url{http://www.bookofspeed.com/chapter2.html}.

\bibitem{Varvello2016}
{\sc Varvello, M., Schomp, K., Naylor, D., Blackburn, J., Finamore, A., and
  Papagiannaki, K.}
\newblock Is the web http/2 yet?
\newblock In {\em Proc. PAM\/} (Crete, Greece, Mar. 2016).

\bibitem{shandian}
{\sc Wang, X.~S., Krishnamurthy, A., and Wetherall, D.}
\newblock Speeding up web page loads with shandian.
\newblock In {\em Proc. USENIX NSDI\/} (Santa Clara, CA, USA, Mar. 2016).

\bibitem{NIPS2010_0577}
{\sc Welinder, P., Branson, S., Perona, P., and Belongie, S.~J.}
\newblock The multidimensional wisdom of crowds.
\newblock In {\em Advances in Neural Information Processing Systems\/} (2010).

\bibitem{Wang2016NSDI}
{\sc Xiao~Sophia, W., Arvind, K., and David, W.}
\newblock Speeding up web page loads with shandian.
\newblock In {\em Proc. USENIX NSDI\/} (Santa Clara, CA, USA, Mar. 2016).

\end{thebibliography}

\end{document}